\documentclass[amsmath,amssymb]{revtex4}

\usepackage{graphicx}
\usepackage{dcolumn}

\usepackage{amsmath,amsthm}
\usepackage{amssymb}
\usepackage{epstopdf}
\usepackage{times}
\usepackage{mathrsfs}
\usepackage{color}
\usepackage{gensymb}
\usepackage{times}
\usepackage{subfigure}
\usepackage[]{graphicx}
\usepackage{amssymb}
\usepackage{amsmath}
\usepackage{bm}
\DeclareGraphicsExtensions{.png,.jpg,.pdf,.gif,.pdf}
\usepackage{amsmath,amssymb,amsthm,amsfonts,eucal}
\usepackage[english]{babel}

\begin{document}

\title{The role of phase compatibility  in martensite}
\author{O\~guz Umut Salman$^{1,2}$}
\author{Alphonse Finel$^{1}$}
\author{Remi Delville$^{3}$}
\author{Dominique  Schryvers$^{3}$}
\affiliation{ $^{1}$Laboratoire d'Etude des Microstructures, ONERA-CNRS, BP 72, 92322  Chatillon, France\\
$^{2}$SEAS, Harvard University, 29 Oxford Street,
Cambridge, MA 02138, USA\\
$^{3}$EMAT, University of Antwerp, Groenenborgerlaan 171, B-2020 Antwerp, Belgium}

\date{\today}
\begin{abstract}
Shape memory alloys inherit their macroscopic properties from their mesoscale microstructure originated from the martensitic phase transformation. In a cubic to orthorhombic transition, a single variant of martensite can have a compatible (exact) interface with the austenite for some special lattice parameters in contrast to conventional austenite/twinned martensite interface with a transition layer. Experimentally, the phase compatibility results in a dramatic drop in thermal hysteresis and gives rise to very stable functional properties over cycling.  Here, we investigate the microstructures observed in Ti$_{50}$Ni$_{50-x}$Pd$_{x}$  alloys  that undergo a cubic to orthorhombic martensitic transformation using a three dimensional phase field approach. We will show that the simulation results are in  very good agreement with transmission electron microscopy observations. However, the understanding of the drop in thermal hysteresis requires the coupling of phase transformation with plastic activity. We will discuss this point within the framework of thermoelasticity, which is a generic feature of the martensitic transformation.
\end{abstract}
\maketitle
\section{Introduction}
Martensitic transformation in shape memory alloys (SMAs) is characterized by its macroscopic reversibility which gives rise to the shape memory effect (SME) or superelasticity \cite{ISI:000227511700001,Bhattacharya:2003qy}. Macroscopic properties of SMAs strongly depend on their microstructures originating from martensitic phase transformation (MT). MT is a first-order displacive transition consisting in a shear-dominated change of the underlying crystal lattice from a high symmetry phase, the austenite to a low symmetry phase, the martensite. The lattice mismatch between the different variants of martensite and the austenite results in long-range elastic interactions. Accommodation mechanisms such as twinning and tapering of martensite at the habit plane with austenite, minimize the strain energy. 
Despite these accommodation mechanisms, high levels of stresses are generated at the interface between the austenite and martensite. Several experimental studies have shown that dislocations were generated at the interface and that they build up during repetitive cycling through the phase transformation \cite{ISI:000274931200036, ISI:000220697800013, ISI:000222122900020, Noreet:2009qr,Delville2011282}. Such dissipative, non-reversible mechanisms are responsible for the significant hysteresis observed in SMAs and the degradation of their functional properties over cycling which eventually leads to failure. The typical approach to tackle this problem is through thermomechanical treatments to strengthen the microstructure (strain hardening \cite{ISI:A1982QE67200035}, nano-structuring \cite{ISI:000259016600033,ISI:000279787100020}, precipitation hardening \cite{ISI:A1982QE67200036,ISI:000178934800017}). 

Recently, however, a different approach has been investigated that simply consists in tuning the lattice parameters of some SMAs to obtain a geometrical fit between austenite and a single variant of martensite (phase compatibility). By doing so, a dramatic drop in hysteresis when phase compatibility is achieved has been observed \cite{ISI:000236530400019, ISI:000269420400001}. Furthermore, in addition to a very small hysteresis, SMAs satisfying the compatibility condition have been shown to have very stable functional properties over cycling \cite{ISI:000279707200006}.

\begin{figure}[ht!]
\begin{minipage}[t]{1.0\linewidth}
\begin{center}
\includegraphics[scale=2]{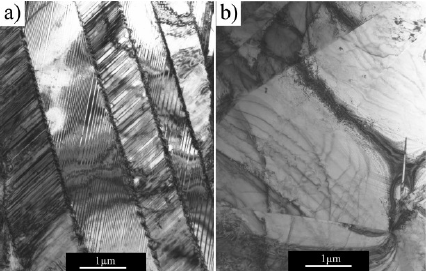}
\end{center}
\end{minipage}
\caption{\label{lambda_diff}\small\textit{TEM study displaying microstructure evolution with respect to Pd composition \cite{ISI:000274582300009,ISI:000262553300007}. (a) Internally twinned martensite plates are observed in $Ti_{50}Ni_{27}Pd_{23}$ alloy with  $\lambda \neq 1$, (b) in contrast, there are large twinless martensite domains in $Ti_{50}Ni_{39}Pd_{11}$ alloy with  $\lambda_2$ very close to 1. }}
\end{figure}

So far, most of the SMAs studied that exhibit phase compatibility undergo a cubic to orthorhombic (c-o) martensitic transformation giving rise to 6 variants of martensite described by their respective transformation stretch matrix $\bold U_i$. A matrix  $\bold U_i$  maps the austenite into one of the martensitic variants and its shape is dictated by the symmetry elements lost by the austenite.    According the geometrically non-linear theory of martensite (GNLTM) (\cite{Bhattacharya:2003qy,ISI:A1992HE58100007}) a compatible interface is possible when the middle eigenvalue $\lambda_2$ of the transformation matrix is equal to 1. In most  alloys this condition is not satisfied and therefore this is referred to as a highly non-generic condition in martensite. An example of a studied c-o system is  Ti$_{50}$Ni$_{50-x}$Pd$_{x}$ whose composition was systematically tuned to achieve geometric compatibility. In addition to a decrease in hysteresis, a dramatic change in the microstructure as one approaches  $\lambda_2=1$  has been reported in reference \cite{ISI:000274576500013}. The transmission electron microscopy (TEM) investigation of microstructures in these alloys shows the presence of large twinless martensite domains in contrast to internally twinned lamellar morphology commonly observed in martensites as illustrated in Figure \ref{lambda_diff}.

Different types of microstructures in alloys that undergo c-o phase transformation have successfully been studied in \cite{Hane:2000uq} by means of  the geometrically non-linear stress-free theories  used for the microstructure optimization as proposed in references \cite{ISI:A1992HE58100007, Ball:1987fr}. It has been shown that one can obtain a very large variety of microstructure in c-o transitions. In a recent work,   the formation of an exact austenite-martensite interface is observed by means of two- dimensional phase field method using  geometrically linear elasticity \cite{ISI:000276554600030}.

The goal of the present work is to carry out simulations of the phase transformation in SMAs using experimental data obtained for  Ti$_{50}$Ni$_{50-x}$Pd$_{x}$ in order to reproduce experimentally observed microstructures. To that purpose, we use a three-dimensional phase field approach that incorporates the strain energy responsible for all specific features of the SMA such as transformation hysteresis and thermoelastic equilibrium.    From there a discussion on hysteresis and thermoelasticity will follow.

\section{Phase field method}
The phase field method is a powerful tool for the investigation of microstructures at mesoscale. It has been successfully used in the study of a large variety of problems in solid-to-solid transformations  \cite{Wang:2004zv,Boussinot2009921,Finel2010245}.  The phase field approach consists in using some continuous  fields of non-conserved and/or conserved parameters. These fields are functions of position and time and they are defined in the entire volume. They assume homogenous values within every phase domain but vary smoothly and continuously through narrow regions that correspond to phase boundaries.   Martensitic phase transformations can be characterized completely by structural non-conserved long-range order parameters  because these transformations are displacive solid state structural changes without atomic diffusion. The temporal evolution of the microstructure can be obtained by solving the usual dissipative time-dependant Ginzburg Landau (TDGL) equations:
\begin{equation}
\label{eq:1}
\frac{\partial\eta_p(\bold r,t)}{\partial t} = -L\frac{\delta \mathscr F_{total}}{\delta \eta_p(\bold r,t)},
\end{equation}
where $L$ is a relaxation coefficient, $\delta $ denotes a functional derivative and $\eta_p$ is the long-range order parameter associated to the martensitic variant $p$.

In order to implement the evolution equations of the order parameters, one has to specify the free energy. The total free energy of a system in a volume V is written as follow
\begin{equation} \mathscr F_{total} =\int_VF_{GL}dV+ \mathscr E_{elastic}.
\end{equation}
In the following, we describe each component of the free energy.

The non-equilibrium chemical free energy of the stress-free crystal is given by a standard Ginzburg-Landau functional
\begin{equation}
\label{eq:totalpf}
F_{GL}(\{\eta_i\}) = F_{L}(\{\eta_i\})  + \sum_i\frac{\beta}{2}|\bold{\nabla}\eta_i|^2,
\end{equation}
where $F_{L} $ is an homogenous  free energy and the second term on the rhs. is a heterogeneous (Ginzburg) term penalizing the spatial variations of the fields.  
The simplest fourth-order polynomial approximation of the Landau free energy density, invariant by permutation of martensitic variants, is given by
\begin{equation}
F_L( \{\eta_i \}) = A\sum_i\eta_i^2 - B\sum_i\eta_i^3+ C(\sum_i\eta_i^2)^2.
\end{equation}
This form may be used if the free energy is not invariant through the symmetry operations $ \eta_i \rightarrow -\eta_i$. The coefficients A, B and C are  positive temperature-dependent constants and can be fitted to reproduce the latent heat of transition, the equilibrium order parameters and the transition temperature.
The gradient term  prevents abrupt changes across the interfaces between different variants of martensitic phases and between the variants and the austenite. Thus, the order parameters change smoothly in a narrow region which means that the interfaces have a finite width in this formulation. Generally speaking the coefficient $\beta$ may be chosen in such a way that the Ginzburg-Landau modeling reproduces some interfacial free energy density and/or interfacial widths.  

An arbitrary microstructure composed of different martensitic variants embedded in the austenite has a residual elastic energy $\mathscr E_{elastic}$ induced by lattice compatibility along the different interfaces. In the framework of linear elasticity, this compatibility-induced elastic energy is given by 
\begin{equation}
\label{eq:elen}
\mathscr E_{elastic}= \frac{1}{2}\int_V C_{ijkl}\bigl(\epsilon_{ij}(\bold{r}) - \epsilon^0_{ij} (\bold{r})\bigr)\bigl(\epsilon_{kl}(\bold{r}) - \epsilon^0_{kl} (\bold{r})\bigr)dV,
\end{equation}
where $C_{ijkl}$ are the elastic moduli~\cite{endnote43}, $\epsilon_{ij}(\bold{r})$ is the actual strain at a given coordinate $\bold{r}$ due to the deformation of the lattice and  $\epsilon^0_{ij}(\bold{r})$ is the local eigenstrain.  Thorough this paper, strain tensors and deformation gradients will be treated within the geometrically linear approximation. The stress-free tensor  $\epsilon^0_{ij}(p)$ of variant p is related to the corresponding transformation stretch matrix  $U(p)$ by
\begin{equation}
\epsilon_{ij}^0(p)= \bold U(p)-\bold I,
\end{equation}
where $\bold I$ is the identity matrix. Within the linear geometry,  compatibility between a martensitic variant and the austenite  is possible when the middle eigenvalue of a stress-free strain tensor is 0.

The actual strain $\epsilon_{ij}(\bold{r})$
can be decomposed into homogeneous and inhomogeneous parts as 
\begin{equation}
\epsilon_{ij}(\bold{r}) =\bar \epsilon_{ij}+   \Delta \epsilon_{ij}(\bold{r}).
\end{equation}
The homogenous strain $\bar \epsilon_{ij}(\bold{r})$ describes the macroscopic shape of the system.  Therefore, the inhomogenous strain $\Delta \epsilon_{ij}(\bold{r})$, which measures local  deviation from the homogenous one,  satisfies 
$
\int_{V}\Delta \epsilon_{ij}(\bold{r})dV=0,
$
and may be related to an inhomogenous  displacement field $\bold u(\bold r)$ by 
$$
\Delta\epsilon_{ij}(\bold r)=\frac{1}{2}(\frac{\partial u_i(\bold r)} {\partial x_j}+\frac{\partial u_j(\bold r)}{\partial x_i}).
$$
Here, we consider the case where the crystal lattice remains coherent upon phase transformation i.e. the total displacement field is continuous.  Furthermore, we assume that the elastic relaxation is much faster than the other relaxation mechanisms so that mechanical equilibrium is reached instantaneously.  Therefore, we have:
 \begin{equation}
 \label{eq:mecheq}
\frac{\delta\mathscr E_{elastic}}{\delta u_i(\bold{r})} = 0.
 \end{equation}
These mechanical equilibrium equations can easily be solved in Fourier space for the displacement field. Inserting the solution back into Eq. \ref{eq:elen}, an explicit representation of the elastic energy follows:   
%
 \begin{equation}
\label{eq:strainlast}
\mathscr E_{elastic} = \frac{V}{2}C_{ijkl}\bar\epsilon_{ij}\bar\epsilon_{kl}
-VC_{ijkl}\bar\epsilon_{ij}\bigl<\epsilon^0_{kl}\bigr>
+\frac{V}{2}C_{ijkl}\bigl<\epsilon^0_{kl}\epsilon^0_{kl}\bigr>
 \frac{1}{2}\int^* n_i\sigma^0_{ij}({\bold q})\Omega_{jk}(\bold n)\sigma^0_{kl}(\bold q)n_l\frac{d\bold q}{(2\pi)^3},
 \end{equation}
where  $\hat X(\bold{q}) = \frac{1}{V}\int_V X(\bold{r}) e^{-i\bold{q}. \bold{r}}d\bold{r} $ is the Fourier transform of $X$, $\bold q$ is the wave vector,  $\bold{n} = \frac{\bold{q}}{|\bold{q}|}$ and   $\bigl<f(\bold r)\bigr> =\frac{1}{V}\int_Vf(\bold r)d\bold r$. The tensor $\Omega_{ik}(\bold n)$ is the Green tensor of the linear elasticity defined by $\Omega_{ik}^{-1}(\bold n) =C_{ijkl}n_jn_l $.   The notations $\int^*$ in Eq. \ref{eq:strainlast} refers to the part $\bold q=0$ that is excluded from the integration domain. The strain energy (\ref{eq:strainlast}) depends on macroscopic deformation, which itself depends on the choice of boundary conditions. Thorough  this paper, we adopt clamped boundary condition $\bar\epsilon_{ij}=0$, which is a convenient choice since in most of SMAs the macroscopic deformation is zero.
 
In all generality, the local eigenstrain $\epsilon_{ij}^0(\bold{r})$  that enters into the elastic energy $\mathscr E_{elastic}$ must be related to the long-range order parameters $\eta_p(\bold r)$   that enter into the Ginzburg-Landau functional . To the lowest order, and as here the total free energy is not invariant by the transformation $\eta_p \rightarrow -\eta_p$, we have: 
\begin{equation}
\label{eq:sfrel}
\epsilon^0_{ij}(\bold{r}) = \sum_{p=1}^{n_{order}}\epsilon^{0}_{ij}(p)\eta_p(\bold{r}),
\end{equation}
Using the above relation, the strain energy may be expressed explicitly  in terms of order parameters and it is given by 
\begin{equation}
\label{eq:fambpq}
\mathscr E_{elastic} =\frac{1}{2}\sum_{p,q=1}^{n_{order}} \int_VB_{pq}(\bold{q})\eta_p(\bold{q})\eta_q^*(\bold{q})\frac{d\bold q}{(2\pi)^3},
\end{equation}
The matrices $B_{pq}$ are defined as
\begin{equation}
\label{eq:Bpq}
B_{pq}(\bold{q}) =  C_{ijkl}\epsilon_{ij}^0(p)\epsilon_{kl}^0({q}) - n_i\sigma^0_{ij}(p)\Omega_{jk}(\bold{n})\sigma^0_{kl}(q)n_l,
\end{equation}
where  $\sigma^0_{ij}(p)$ is defined by $\sigma^0_{ij}(p)=C_{ijkl}\epsilon_{kl}^0(p)$. The matrices $B_{pq}(\bold q)$ characterize the long range  elastic interactions between different phases. In  equation (\ref{eq:fambpq}),  the notation $X^*$ is the complex conjugate of the quantity $X$. 
\section{Numerical study}
Our aim here is to perform dynamical simulations using the phase field model  to understand the influence of  the middle eigenvalue  $\lambda_2$ on the microstructure.  We perform simulations corresponding to identical physical systems in terms of thermodynamical potential and interfacial energy but that differ  in terms of lattice parameters . We have chosen to use two TiNiPd alloys, namely Ti$_{50}$Ni$_{50-x}$Pd$_{x}$ with $x=0.11$ and $x=0.25$, for which $\lambda_2=1$ and  $\lambda_2=1.018$ respectively,  to monitor the effect of the value of $\lambda_2$ on microstructure evolution (see reference \cite{ISI:000262553300007} for lattice parameters). Additionally, we repeat the simulations using the lattice parameters of TiNiCu, satisfying the $\lambda_2=1$ condition but with an important volumetric strain  ($Tr(\epsilon_{ij}^0) \approx 0.02$)   associated with the martensitic transformation. The volumetric strains of   TiNiPd alloys are much smaller i.e. $Tr(\epsilon_{ij}^0)  \approx 0.004$.     
\subsection{Simulation parameters}
The input parameters of the phase field method are the elastic constants, the transformation strains, the interfacial energy and free energy functional.  To make sure that microstructural evolutions will depend only on the value of $\lambda_2$ for the three alloys in question, all simulation parameters are chosen to be same, except the transformation strains ($\epsilon_{ij}^0$).

The total  energy density in a reduced form is written  as
\begin{equation}
\label{eq:landau_c5}
\tilde F( \{\eta_p \}) = \tilde F_{L}+ \frac{\tilde\beta}{2}|\bold{\tilde\nabla}\eta_p|^2 + \tilde E_{elastic},        
\end{equation}  
where $\tilde F_{L} = \frac{F_{L}}{f_0}$, $\tilde\beta = \frac{\beta}{f_0d_0^2}$, $\tilde\nabla =d_o\nabla$,  $\tilde E_{elastic}=\xi \frac{E_{elastic}}{C_{44}\epsilon_0^2}$ with $\xi=\frac{C_{44}\epsilon_0^2}{f_0}$.  Here, $f_0$ is a typical  energy density,  $\epsilon_0$ is a typical strain, $d_0$ is a grid size. The  coefficient $\xi$ is the ratio of  typical strain energy to the driving force.

In the simulations, we choose the following coefficients: $\tilde A=0.2$, $\tilde B=3$ and $\tilde C=2$, $\tilde\beta =0.5$ and $\xi=1$, with these values, the order parameter $\eta_p$ varies from zero in the austenite to 1 in the martensite. The typical strains are  $\epsilon_0 = 0.108$  for Ti$_{50}$Ni$_{39}$Pd$_{11}$, $\epsilon_0 = 0.124$ for Ti$_{50}$Ni$_{27}$Pd$_{23}$ and $\epsilon_0 = 0.0735$ for TiNiCu. For each alloy, we use the same elastic moduli: $C_{11}=142\rm{GPa}$, $C_{12}=125\rm{GPa}$ and $C_{12}=95\rm{GPa}$.  The grid size of the simulation can be found using the relation between the experimental interfacial energy $\sigma$ 
 and the dimensionless interfacial energy $\tilde\sigma$ 
\begin{equation}
\label{rel_int_c4} 
\sigma= \tilde\sigma f_0d_0.
\end{equation}
 One can introduce an interface between two variants and then allows the system to relax to its equilibrium without elasticity. The dimensionless interfacial energy $\tilde\sigma$ of such a  configuration is given by       
\begin{align}
\begin{split}
\label{eq:num_ener_c4}
\tilde\sigma = \sum_r\biggl\{\tilde F_L - \tilde F_{ref} + \sum_p\frac{\tilde\beta}{2}\nabla|\eta_p|^2\biggr\},
\end{split}
\end{align}
where $\tilde F_{ref}$ is the free energy of  a single variant reference state (without interfaces). 
Using Eq. \ref{rel_int_c4}, $d_0$ is found to be
\begin{equation} 
\label{gridsize}
d_0 = \frac{\xi\sigma}{\tilde\sigma C_{44}\epsilon_0^2}.
\end{equation}
Using the parametrization presented above, the grid size $d_0$ is calculated to be   $17\rm{nm}$, $21\rm{nm}$ and $49\rm{nm}$ for the alloys Ti$_{50}$Ni$_{39}$Pd$_{11}$,  Ti$_{50}$Ni$_{27}$Pd$_{23}$ and  TiNiCu, respectively. The required experimental interfacial energy for this calculation was taken from \cite{Artemev:2001ta} due to lack of specific  experimental data for the alloys investigated here. 

The kinetic equations Eq. \ref{eq:1} are integrated in parallel, using a semi-implicit Euler scheme in reciprocal space \cite{Muite}. Also, the choice of the numerical value of the stifness parameter $\tilde \beta$ ensures that the interfaces are large enough to avoid pinning on the numerical grid.
\subsubsection*{Initial conditions}
The martensitic phase transformation has an athermal character. This means that thermal fluctuations do not play any major role. In general, the nucleation occurs around crystal defects \cite{Zhang:2007sf} that decrease elastic energy barriers; whereas  the forces due to thermal fluctuations are small compared to mechanical forces. Dislocations can be incorporated in the phase field model using the equivalence between the elastic fields created by a dislocation loop and those of a platelet inclusion  \cite{Rodney:2001lr,Wang20011847,finelrod}.   A single dislocation was put in the middle of the system and all order parameters $\{\eta_i\}$ were set at zero at time $t=0$. The dislocation loop lies at the $yz$ plane and has a shape of a square with 3 grid points along every edge.  Its normal vector is given by $\bold{\hat n} = (1,0,0)$ and the Burgers vector is chosen to be $\bold{\hat b} = (0,0,1)$.  The simulation results are presented in the following.
\subsubsection*{Microstructural evolution in Ti$_{50}$Ni$_{39}$Pd$_{11}$  }
\begin{figure}[ht!]
\begin{center}
\subfigure[$\bold{t=23\times10^3}$]{\label{fig_11c}\includegraphics[scale=0.4]{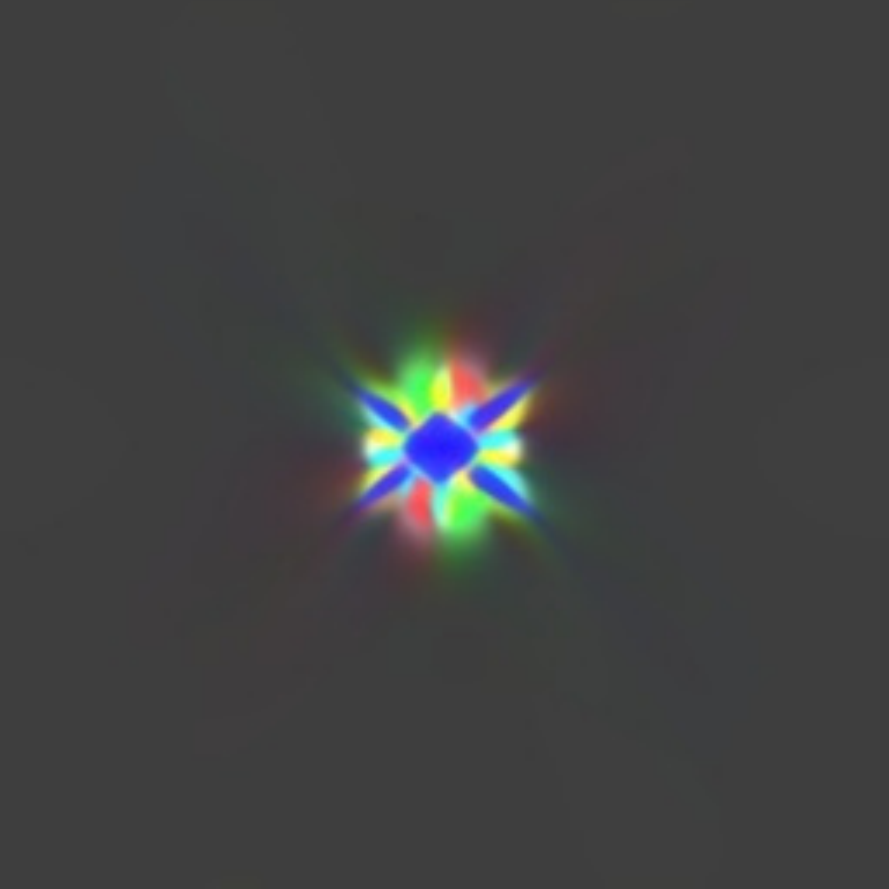}}\hspace{.01in}
\subfigure[$\bold{t=29\times10^3}$]{\label{fig_11d}\includegraphics[scale=0.4]{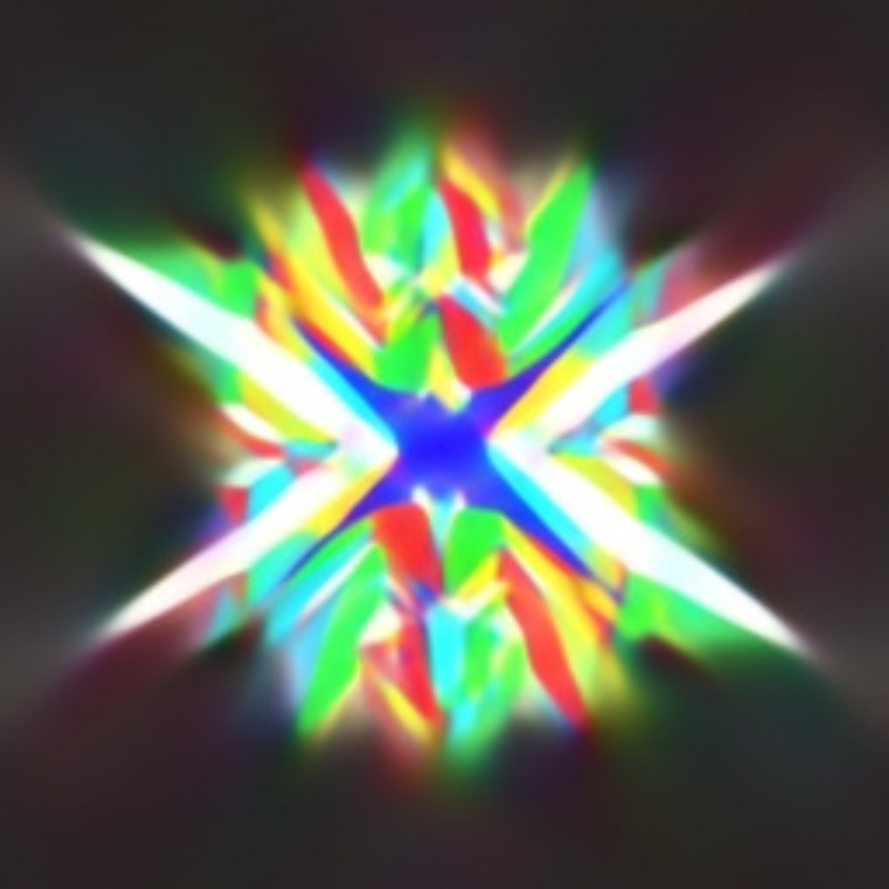}}
\subfigure[$\bold{t=30\times10^3}$]{\label{fig_11e}\includegraphics[scale=0.4]{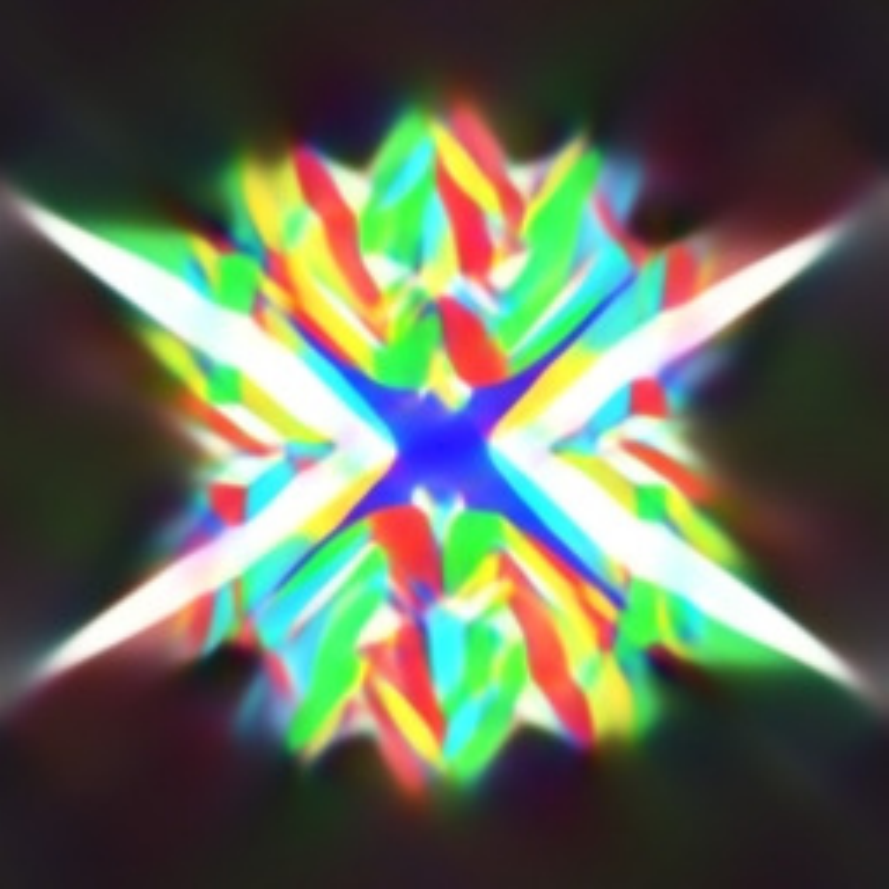}}\hspace{.01in}
\subfigure[$\bold{t=40\times10^3}$]{\label{fig_11f}\includegraphics[scale=0.4]{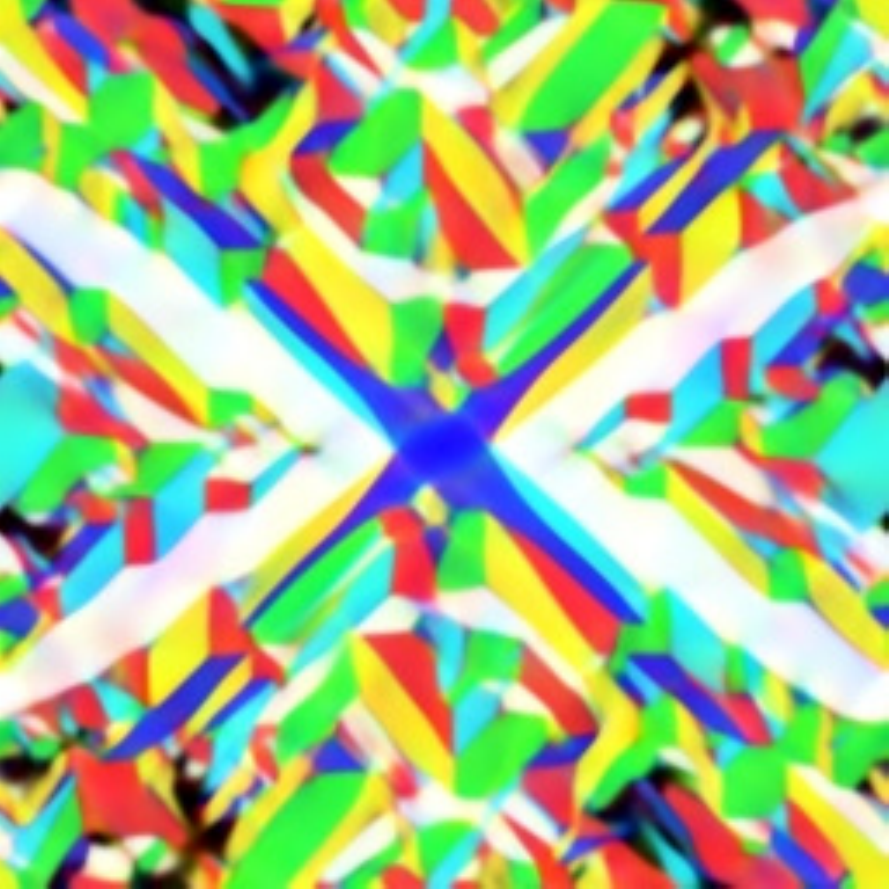}}\\
\end{center}
\caption{\label{fig_11}\small\textit{Time evolution of the microstructure for Ti$_{50}$Ni$_{39}$Pd$_{11}$ alloy, for which $\lambda_2=1$. The simulation box size is $256\times256\times256$.}}
\end{figure}

\begin{figure} [ht!]
\begin{center}
\subfigure[]{\label{pd11simexpb}\includegraphics[scale=0.18]{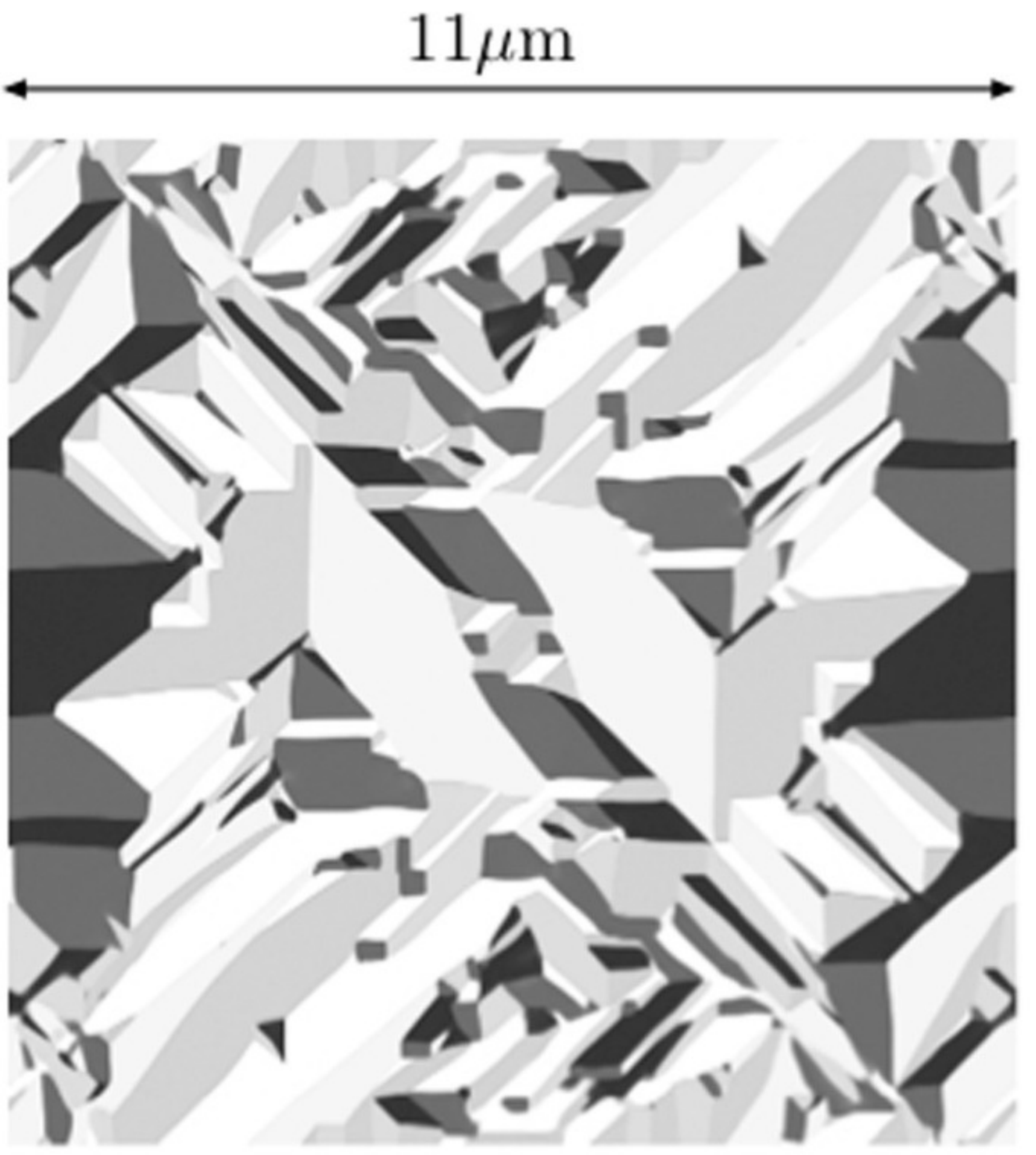}}
\subfigure[]{\label{pd11simexpa}\includegraphics[scale=0.18]{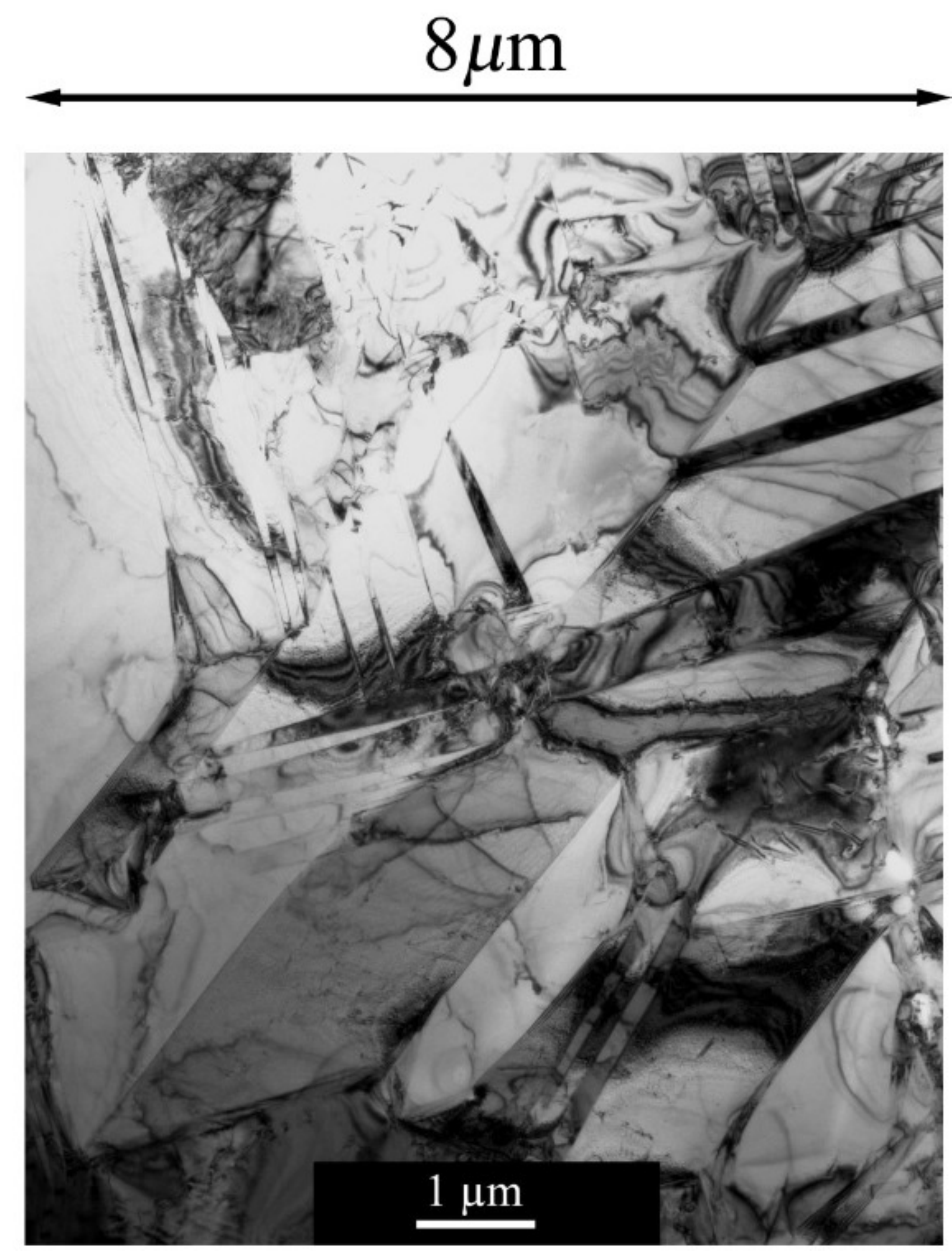}}
\end{center}
\caption{\label{pd11simexp}\small\textit{Comparison of the simulated microstructure with experimental observations \cite{doi:10.1080/14786430903074755}. In both cases, $\lambda_2=1$ condition is satisfied.}}
\end{figure}

Figure \ref{fig_11} shows the microstructural evolution in Ti$_{50}$Ni$_{39}$Pd$_{11}$ alloy. The figures correspond to the middle plane perpendicular to the   $z$ direction of a $256\times256\times256$ simulation box.  The temperature was held constant during the entire evolution.  The austenite (black) transforms into variants of martensite around the dislocation loop. All six variants appear simultaneously (variants 1, 2, 3, 4, 5 and  6 are shown by red, green, blue, white, yellow, light blue, respectively). It is important to point out that we do not observe the growth of  any  typical polytwinned plates formed by   pair of variants with specific twin ratios. Rather, we observe a chaotic microstructure formed by all twin-related variants with inequivalent  volume fractions. For example, it can be seen from figs. \ref{fig_11}\subref{fig_11d}-\subref{fig_11e} that there is a rather large region formed by variant 4 (white), while other variants occupy smaller volume fractions. The simulated microstructure is qualitatively similar to the ones observed experimentally, as seen in Fig. \ref{pd11simexp}, where we present a comparison between simulation and observation.   It is clear that variants with large or small volume fractions are present in the crystal as well as  in the simulated microstructure. 

The origin of this chaotic microstructure is essentially due to the fact that the $\lambda_2=1$ condition allows for the growth of monovariant domain within the austenite without the need of a twinning mechanism to relax stress accumulation along the martensite/austenite interface. The transformation process may be qualitatively described as follows.

As mentioned previously, the nucleation occurs on single defects, here a simple dislocation loop. The first embryo is inhomogenous, but does not display any regular stacking of twins. Its morphology and initial growth are not controlled by compatibility relation between martensite and austenite, nor by the twin relations between martensitic variants, but by the stress-field of the pre-existing defects (see Fig. \ref{fig_11c}). At this stage, the interfaces between the embryo and the austenite do not necessarily fulfill the compatibility relation. Any component of the embryo interface which is non-compatible will favor, by a local twinning process, the subsequent growth of other variants. If, however, it happens that at some point of the formation process, a variant displays a nearly compatible interface with the austenite, as variant 4 (white) in Fig.  \ref{fig_11d}, it may adopt a "pre-needle" shape composed of two nearly parallel and stress-free interfaces that terminate on a rounded tip. In front of the tip, the stress in the austenite will be high and, therefore, the transformation driving force much higher than along the two nearly compatible sides of the needle. This will favor the propagation of the tip and therefore, the rapid growth of the needle-shaped martensite, as seen in Figs.   \ref{fig_11c}-\ref{fig_11f}, where we observe that  variant 4 rapidly reaches the system boundaries, and therefore,  occupies a larger region than the other variants  in the final microstructure.  

In order to understand the absence of laminates in the microstructure, we  checked  that, when the $\lambda_2=1$ condition is fulfilled, the residual elastic energy associated to a polytwinned domain formed by two different variants is always non-zero if the volume fraction of the variants are finite (see Appendix). This analysis      shows that  a twinned-martensite/austenite interface will be energetically non favorable when  $\lambda_2=1$. Hence, a polytwinned domain cannot grow within the austenite.

\begin{figure}[ht!]
\begin{center}
\subfigure[$\bold{t=55\times10^3}$]{\label{finalstates11_25a}\includegraphics[scale=0.11]{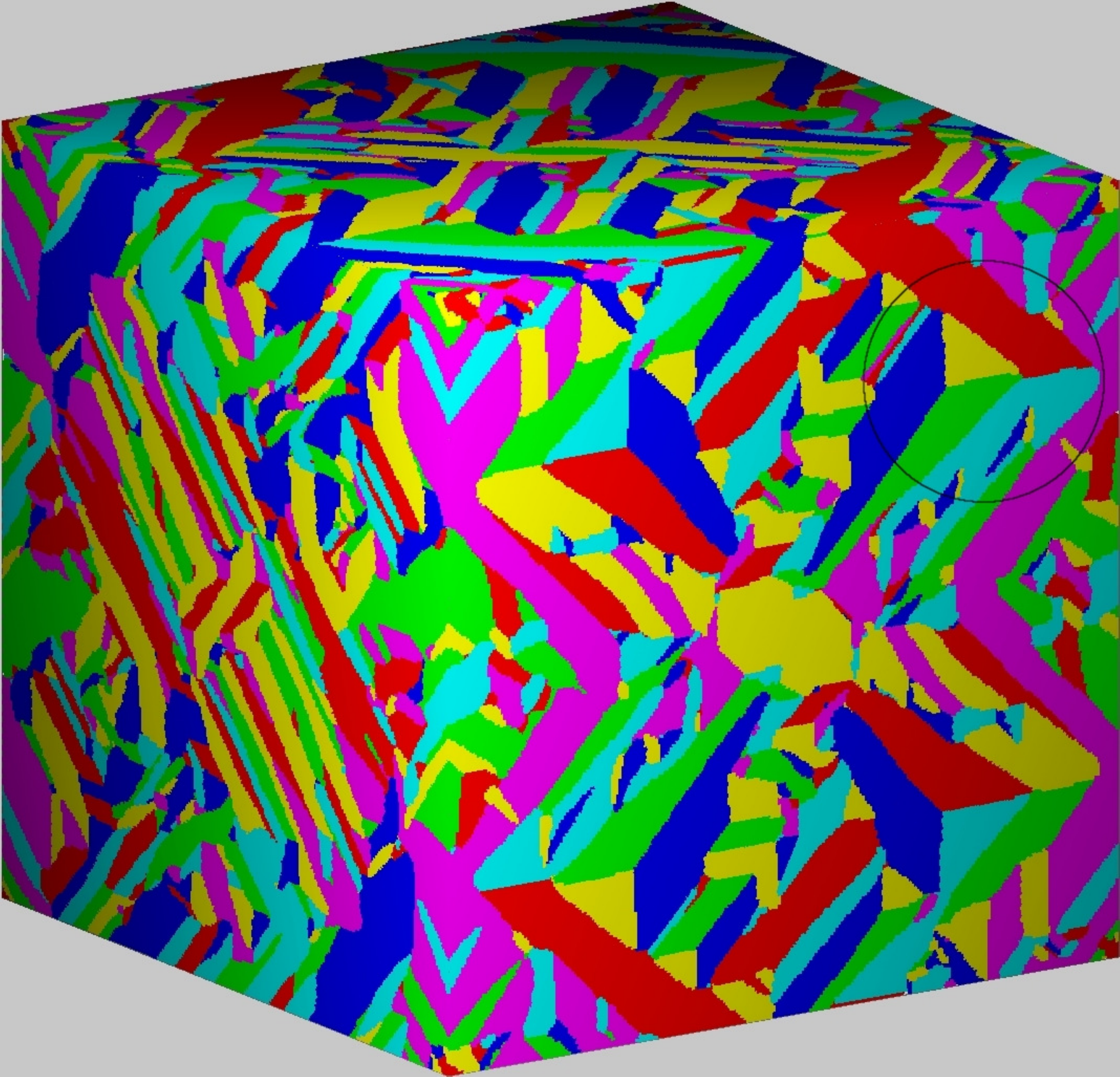}}\hspace{.1cm}
\subfigure[$\bold{t=168\times10^3}$]{\label{finalstates11_25b}\includegraphics[scale=0.11]{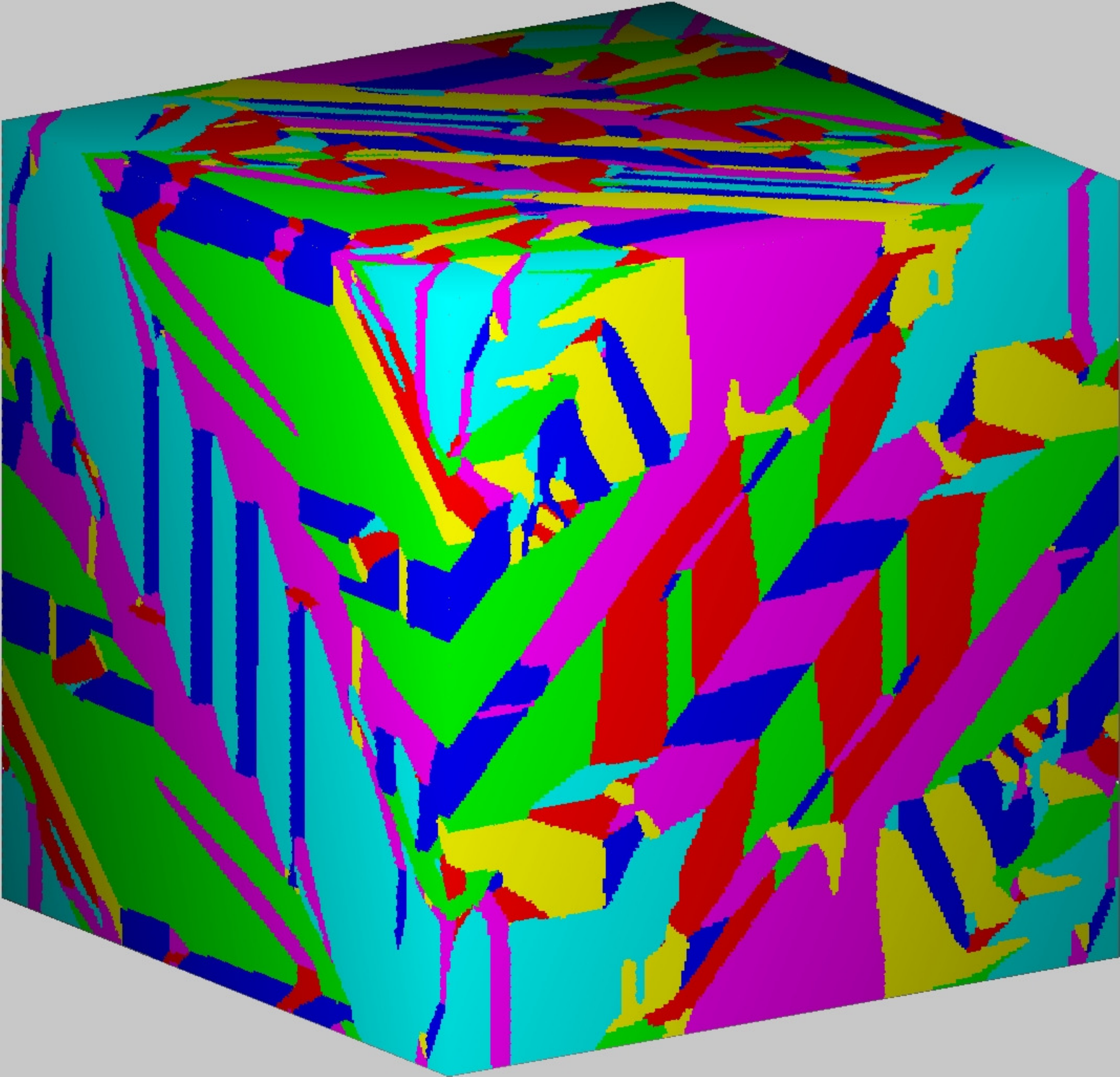}}
\end{center}
\caption{\label{finalstates11_25}\small\textit {Comparison of the final states of the alloys with  different $\lambda_2$. (a) Final state of Ti$_{50}$Ni$_{39}$Pd$_{11}$ ($\lambda_2=1$), and (b) \,\,Ti$_{50}$Ni$_{27}$Pd$_{23}$ ($\lambda_2\neq1$).}}
\end{figure}

\begin{figure} [ht!]
\begin{center}
\subfigure[]{\label{self_accoma}\includegraphics[scale=0.195]{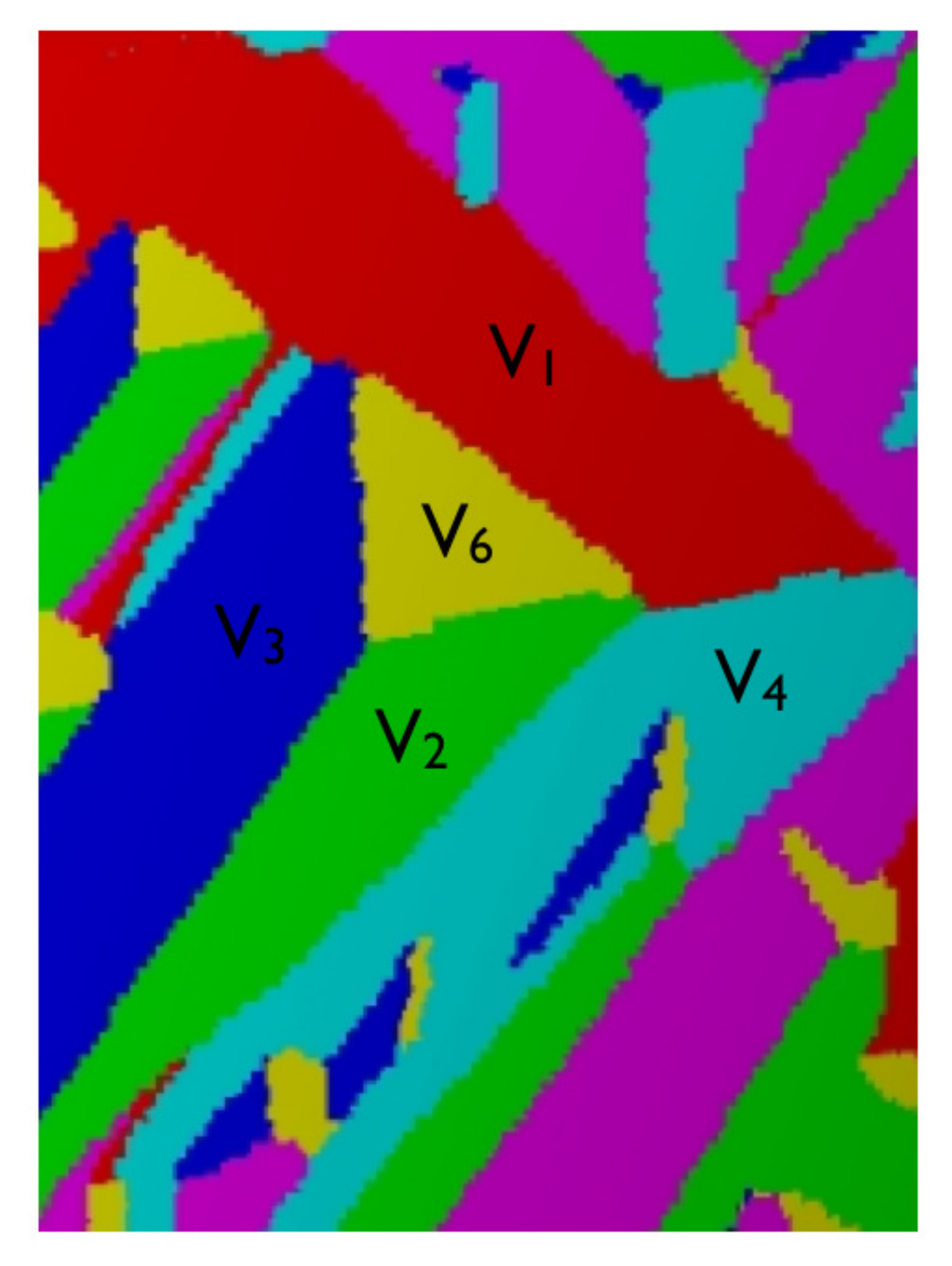}}\hspace{.1cm}
\subfigure[]{\label{self_accomb}\includegraphics[height=4.49cm]{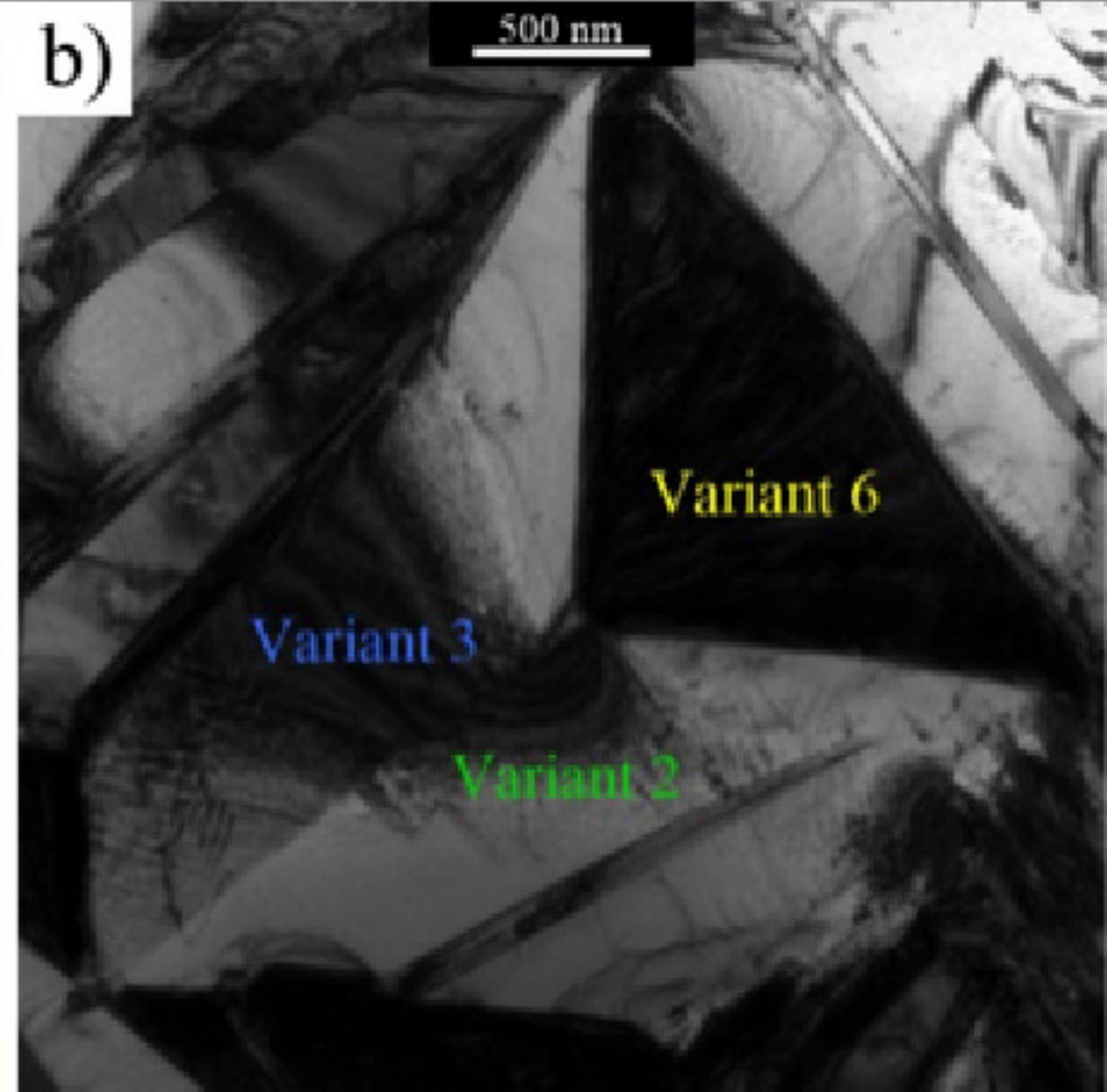}}
\end{center}
\caption{\label{self_accom}\small\textit {(a) Magnification of the encircled region in Fig. \ref{finalstates11_25}\subref{finalstates11_25a}. It shows a typical self-accommodation unit formed by variants 6-3-2.  (b) TEM image displaying the microstructure of Ti$_{50}$Ni$_{39}$Pd$_{11}$ alloy. It exhibits a triangular self-accommodation unit similar to the simulation results.}}
\end{figure}

The three dimensional final state of this simulation is shown in Fig. \ref{finalstates11_25}\subref{finalstates11_25a}. Another microstructure of interest identified in this figure is the triangular structure (encircled region in Fig. \ref{finalstates11_25}\subref{finalstates11_25a}) formed by three variants. The magnification of this structure is shown in Fig. \ref{self_accom}\subref{self_accoma}. Similarly, one can see another self-accommodation unit in Fig. \ref{finalstates11_25}\subref{finalstates11_25a} formed by variants 4-1-3 (white-red-blue). This kind of microstructure was also observed in the experiments when compatibility condition is fulfilled, as seen in Fig. \ref{self_accom}\subref{self_accomb}.

In brief, we understand that the martensitic transformation may proceed by using simultaneously (along different portions of the austenite-martensite interface) or alternatively two mechanisms:
a "local twinning" mechanism as well as the "compatible" one. The additional freedom afforded by the later is at the origin of the chaotic morphology developed by the martensite, at the expense of the more regular polytwinned domains that would be observed if the twinning mechanism was operating alone.
\subsubsection*{Microstructural evolution in Ti$_{50}$Ni$_{27}$Pd$_{23}$  }
\begin{figure}
\begin{center}
\subfigure[$\bold{t=23\times10^3}$]{\label{fig_pd23c}\includegraphics[scale=0.4]{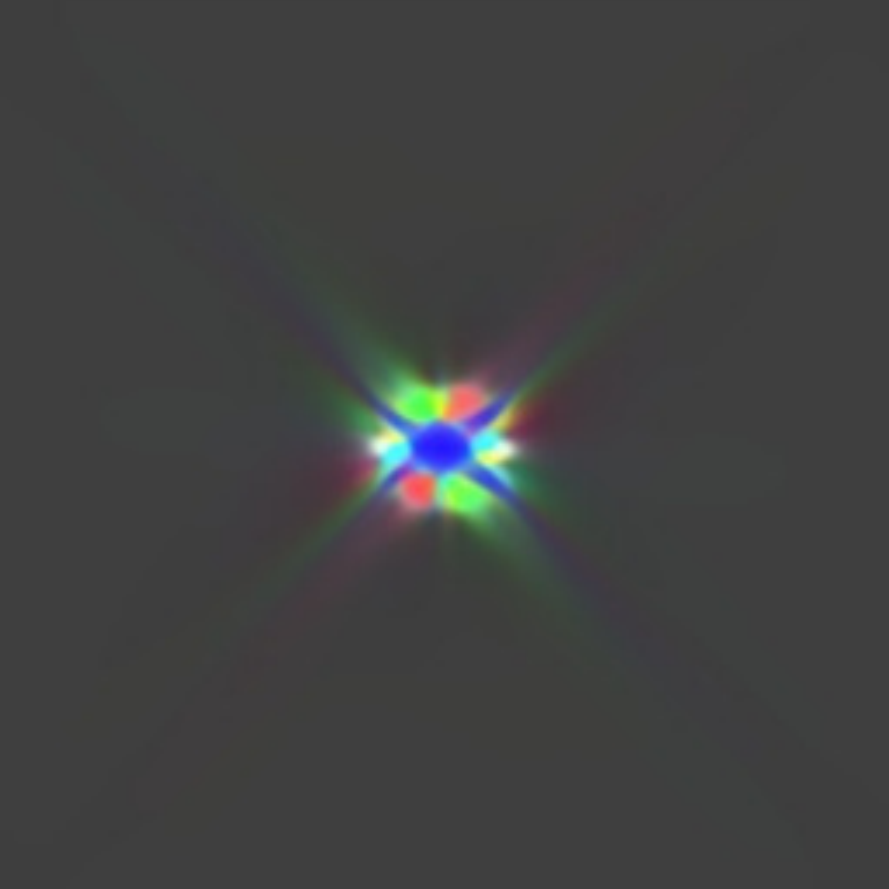}}\hspace{.05in}
\subfigure[$\bold{t=29\times10^3}$]{\label{fig_pd23d}\includegraphics[scale=0.4]{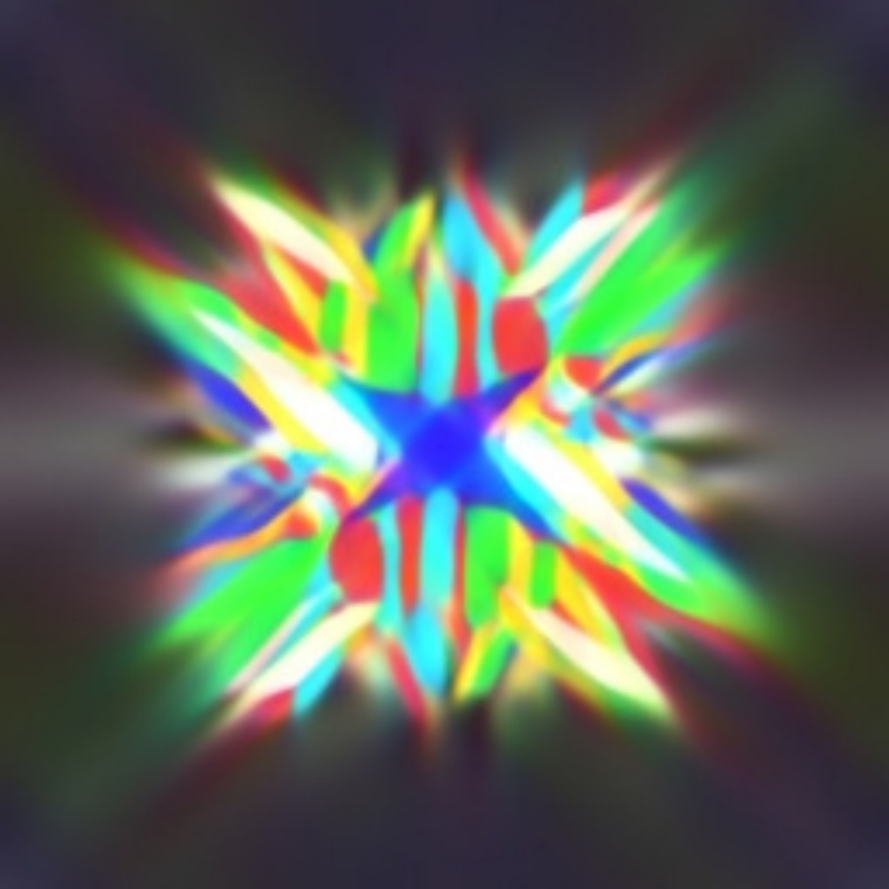}}
\subfigure[$\bold{t=30\times10^3}$]{\label{fig_pd23e}\includegraphics[scale=0.4]{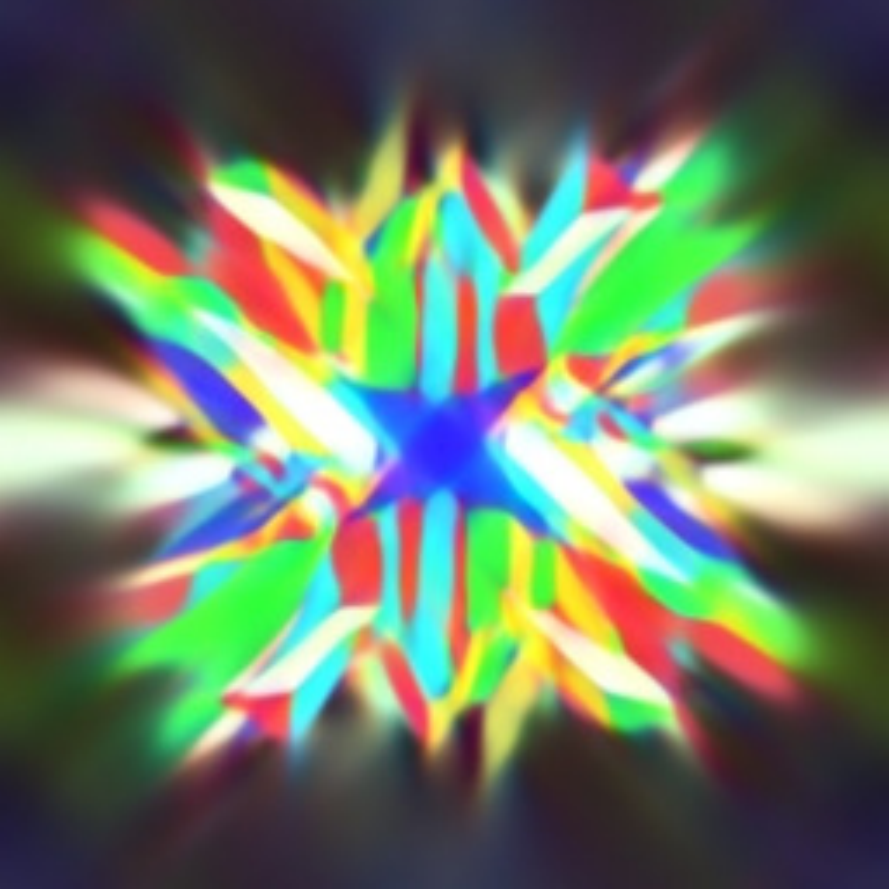}}\hspace{.05in}
\subfigure[$\bold{t=40\times10^3}$]{\label{fig_pd23f}\includegraphics[scale=0.4]{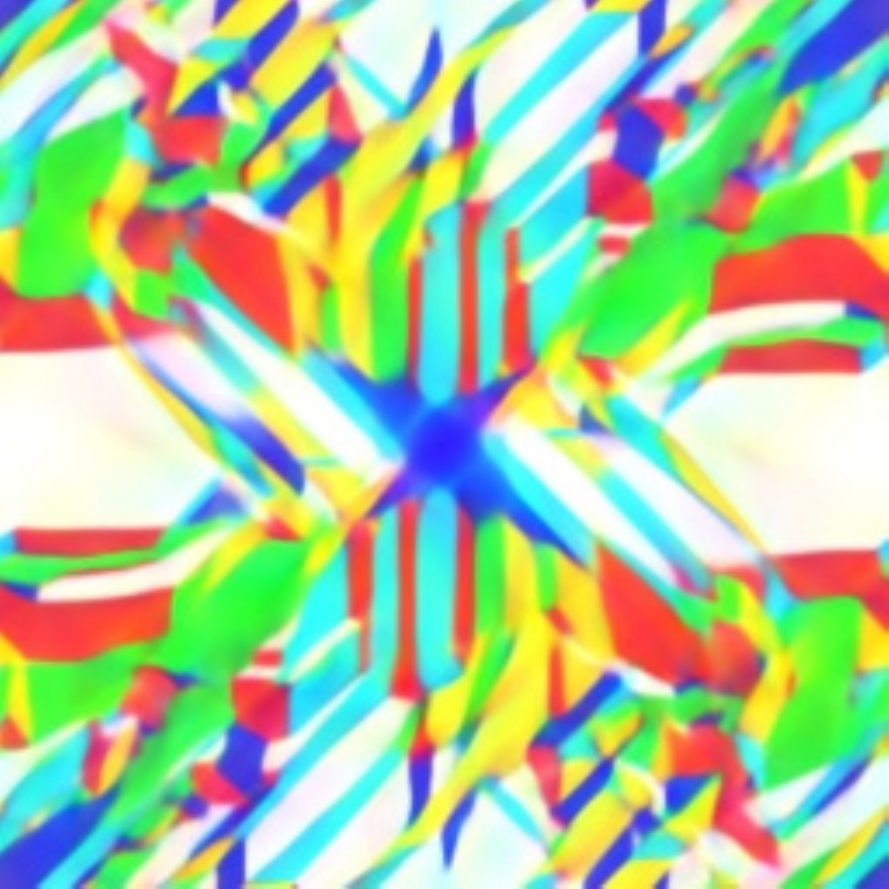}}\hspace{.05in}
\end{center}
\caption{\label{fig_pd23}\small\textit{Time evolution of the microstructure for Ti$_{50}$Ni$_{27}$Pd$_{23}$ alloy, for which $\lambda_2
\neq1$.  The simulation box size is $256\times256\times256$.}}
\end{figure}

Figure \ref{fig_pd23} shows the microstructural evolution in Ti$_{50}$Ni$_{27}$Pd$_{23}$ alloy. We use the same  initial conditions  as in previous simulation. As above, the first embryo is inhomogenous and its initial growth is controlled by the stress-field of the dislocation (see Fig.\ref{fig_pd23c}). After this initial growth,  polytwinned domains compatible with the austenite appear. For instance, variant 1 (red)-variant 5 (light blue), variant 6 (yellow)-variant 2 (green) and variant 6 (yellow)-variant 3 (blue) form laminates that grow into the austenite as shown in Figs. \ref{fig_pd23}\subref{fig_pd23d}-\subref{fig_pd23f}.   The comparison of Fig. \ref{fig_11f} and Fig. \ref{finalstates11_25a}, that correspond to a situation with $\lambda_2=1$, with  Fig. \ref{fig_pd23f} 
and Fig. \ref{finalstates11_25b} , for which  $\lambda_2=1.018$, shows that the microstructure evolution is very sensitive to the value of the middle eigenvalue  $\lambda_2$: a slight deviation from compatibility generates a microstructure which is  significantly more regular than the one observed if  $\lambda_2=1$. Finally, in Fig. \ref{finalstates11_25}\subref{finalstates11_25b}, we show the three dimensional final state of the microstructure, where we observe several  twinned laminates in contrast to the final state of   Ti$_{50}$Ni$_{39}$Pd$_{11}$   alloy.                
%
%
%
%
%
%
%
\subsubsection*{Microstructural evolution in TiNiCu}
\begin{figure}
\begin{center}
\subfigure[$\bold{t=23\times10^3}$]{\label{tinicuc}\includegraphics[scale=0.4]{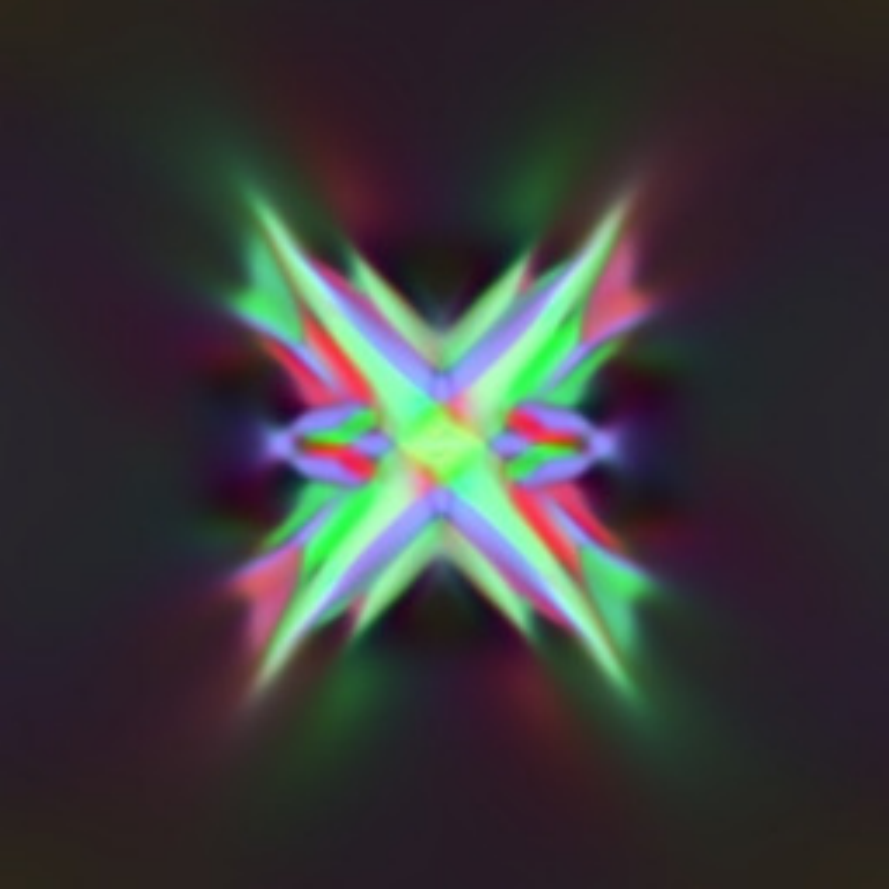}}\hspace{.05in}
\subfigure[$\bold{t=29\times10^3}$]{\label{tinicud}\includegraphics[scale=0.4]{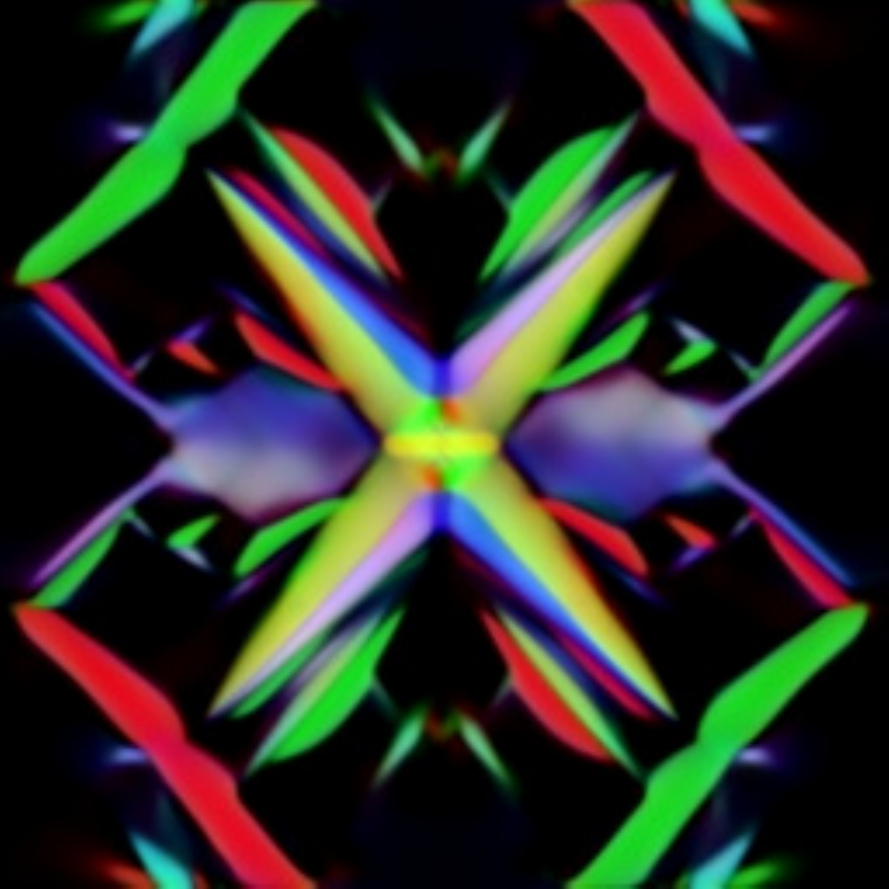}}
\subfigure[$\bold{t=45\times10^3}$]{\label{tinicug}\includegraphics[scale=0.4]{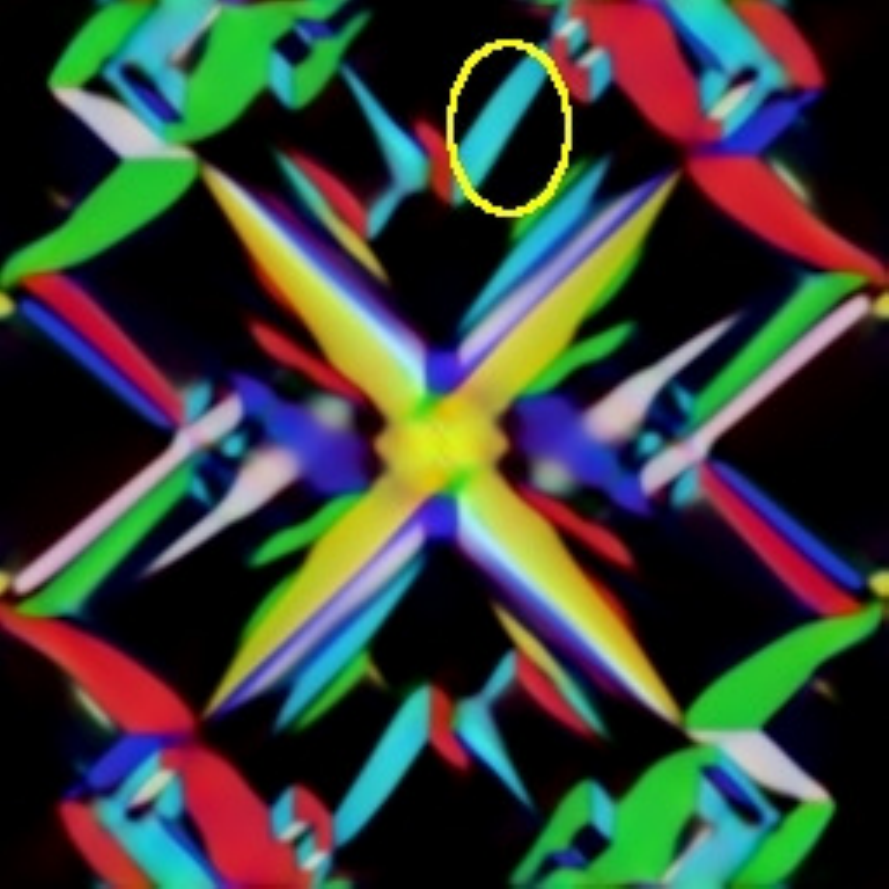}}\hspace{.05in}
\subfigure[$\bold{t=168\times10^3}$]{\label{tinicuh}\includegraphics[scale=0.4]{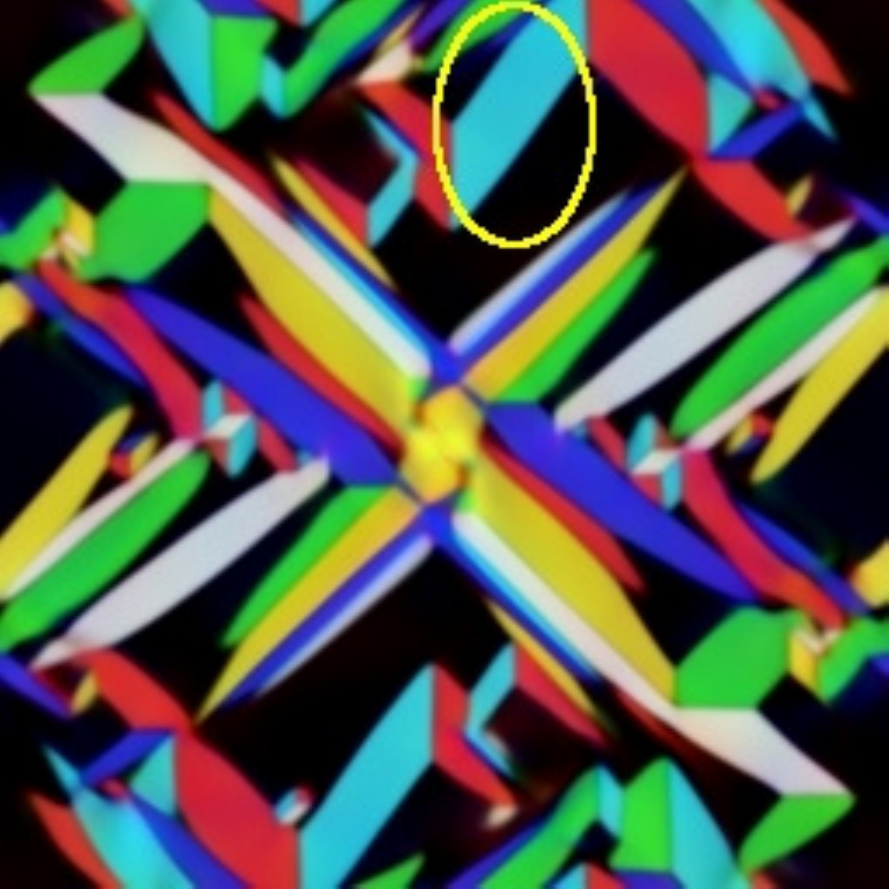}}\hspace{.05in}
\end{center}
\caption{\label{tinicu}\small\textit{Time evolution of the microstructure for TiNiCu alloy, for which $\lambda_2=1$ but the volume change associated with the transformation is non-null. The simulation box size is $256\times256\times256$.}}
\end{figure}

The microstructural evolution in  TiNiCu alloy satisfying the condition $\lambda_2=1$ but with an important volumetric  strain in the martensitic phase subjected to  same initial conditions as in previous examples is shown in Fig. \ref{tinicu}. However, the temperature was decreased continuously. The reason for this is that the presence of the large volume effect causes the transformation to reach its thermoelastic-equilibrium. As previously, in the early stage, the complex multi-variant structure grows into the austenite around the defect. But even at late times, there is retained austenite in the system. We observe that variants form two nearly parallel compatible interface with the austenite that terminate on a tip. As explained previously, in front of the tip, the stress in the austenite is high, and thus, it  results in higher transformation driving force. Therefore, mono-variant domains grow  perpendicular  to the habit plane normal direction. When a variant tip reaches the system boundaries and/or meets another martensitic domain, the further growth in front of the tip is not  possible. At this stage, we  observe that variants grow in the direction of the habit plane normal as shown in Figs. \ref{tinicu}\subref{tinicug}-\subref{tinicuh} in the encircled region. There is no self-accommodation by twinning as observed experimentally \cite{ALLOYS:1990fk}. For this alloy, microstructural evolution is affected by both the large volume effect and the value of $\lambda_2$.  
%
%
%
%
%

%
%
%
%
%
\section{Thermoelasticity}
The phase field simulations presented above reproduce microstructures in good agreement with experimentally  observed microstructures.  However, when we calculated the volume fraction of martensite with respect to the temperature, we did not find a significant influence of the $\lambda_2$ value on the transformation hysteresis. In this section, we discuss this contradiction by considering the thermodynamics of thermoelastic martensitic transformations.

Thermoelasticity is a generic feature of the martensitic transition in SMAs. Basically, it manifests itself by the fact that the martensitic transition spreads over a finite temperature range, generally with a small thermal hysteresis. The forward transition (from austenite to martensite) starts at a temperature $M_s$ and generally finishes at a lower temperature $M_f$. The reverse transformation on heating starts at a temperature $A_s$ and in general comes to completion at a higher temperature $A_f$ with $M_f<A_s$ and $M_s<A_f$. The width of the hysteresis may be quantitatively characterized by $|M_s-A_f|$ and $|M_f-A_s|$ and its spread by  $|M_f-M_s|$ and $|A_s-A_f|$ (see Fig. \ref{fig:hysteresisthermob}). 

In principle, in the context of equilibrium thermodynamics, the martensitic transition should take place  at the temperature $T_0$ for which the stress-free (or ¬chemical¬) Gibbs energies of the austenite and martensite, given by $G_M=H_M-TS_M$ and $G_A=H_A-TS_A$ (i.e., the free enthalpies), are equal. 

If the transition did take place precisely at the equilibrium temperature $T_0$, it would be \textit{completly reversible}, with no hysteresis and accompanied by a latent heat $\Delta Q$ (measured by calorimetry) equal to the difference between the austenite and the martensite enthalpies: $\Delta Q = H_A(T_0)-H_M(T_0)$. 

In reality, the situation is more complex and, as mentioned above, the forward and reverse transformations spread over a finite temperature range and display a (small) hysteresis. A large quantity of experimental and theoretical studies have been devoted to understand the origin of those two phenomena   (see \cite{Wollants:1993rz}, \cite{Ortin:1989rm} and \cite{oplanes} for reviews).

However, there is still often some confusion concerning the correct comprehension of these features, even though it is clearly understood that the origin of thermoelasticity and of the finite width of the hysteresis are due to stored elastic energy and frictional work. 

The underlying physical mechanisms responsible for the spreading of the forward and reverse transitions and of the hysteresis are qualitatively described in the following. We in particular clearly explain that thermoelasticity by itself is a thermodynamically reversible process (in other words, the process of storing elastic energy is not irreversible), whereas the presence of local free energy barriers (such as those due to friction on local external defects or to the competitive growth of different martensitic variants) leads to irreversibility and intermittent dynamics. This analysis is illustrated through the discussion of two different simple situations that involve the interplay between stored elastic energy and a local constraint (the model of Olson and Cohen) or fixed boundary conditions (a clamped system).
%
\subsubsection*{The martensitic transition in a perfect single crystal} 
Due to the large strain associated to the martensitic transition, nucleation of a martensitic domain requires accommodation by twinning. Therefore, the first embryos are essentially modulated or simply sheared as onset for the polytwinned character of the final martensite domain,  and cannot nucleate homogeneously through thermal fluctuations \cite{PhysRevB.44.9301,Murakami1}. Nucleation is heterogeneous on pre-existing defects that generate local favorable strains. Note that, to a certain extent, these embryos exist above the equilibrium temperature $T_0$ and can be responsible for the softness of the effective shear elastic constant often observed experimentally well above $T_0$. From this point of view, the "soft mode" phenomenon often referred to as pre-martensitic and cited to be a characteristic of the martensitic transition, is not an intrinsic softening of the austenite phase, but rather a collective and effective response of pre-existing martensitic-like embryos stabilized by static defects, such as dislocations and/or grain boundaries or even pinned defects \cite{PhysRevB.44.9301,Murakami1,endnote44}

Inside the martensitic embryos accommodation cannot be perfect, because the boundary of any finite domain cannot consist exclusively of habit planes. Therefore, elastic screening by twinning is only partial and some unrelaxed elastic energy is stored in the resulting martensitic domain and in its surrounding. Below $T_0$, when $G_M(T)$, the Gibbs free energy of martensite is lower than $G_A(T)$, the Gibbs free energy of the austenite, part of the stress-free driving force $\Delta G(T)=G_M(T) - G_A(T)$ is consumed by the stored elastic energy. A simple model (see below) shows that, for sufficiently small undercooling, a modulated embryo that nucleates on a defect reaches an equilibrium state with a finite volume. When temperature is sufficiently lowered, this embryo becomes unstable and undergoes a continuous growth at a fixed temperature. In brief, the martensitic start temperature $M_s$ is lower than $T_0$, and corresponds to the first embryo that becomes unstable. $M_s$ is not intrinsic and it depends on the nature of the pre-existing defects, and may change during cycling (because the martensitic transition itself creates defects). The growth of this isolated embryo is \textit{irreversible}, because it occurs at a temperature when the martensite and the austenite are not in thermodynamic equilibrium. Therefore, it is accompanied by an increase of the total entropy (entropy of specimen plus the entropy of the surroundings). If the embryo does not meet any obstacle, the growth is continuous and, in an ideally perfect single crystal, the martensite growth will reach completion at $M_f = M_s$: no thermoelasticity would be observed, even though the growth is continuously accompanied by a stored elastic energy that consumes part of the available \textit{stress-free} driving force. Since the martensitic phase is more stable than the austenite for temperatures smaller than $T_0$, the reverse transformation on heating will start at a temperature $A_s$ strictly above $M_s$.

However, $A_s$ may be below or above $T_0$ depending on the amount of unrelaxed elastic energy stored in the microstructure when completion is reached and the temperature reversed. More precisely,  $A_s$ is the smallest temperature at which, on heating, some martensitic domain becomes unstable with respect to the austenite. If the specimen under consideration is a prefect single crystal and if the forward transition involved only one martensitic plate, it is likely that, at completion, the stored elastic energy was essentially reduced to zero because an almost perfect self-accommodation may be reached when there is no residual austenite and if the external boundaries are free. As a result, the reverse transformation cannot take place before the stress-free bulk austenite become more stable than the stress-free bulk martensite. This implies $A_s>T_0$. Also, as soon as the reverse transition starts, it will go to completion at fixed temperature, because all the components of the total Gibbs free energy operate in the same direction (the free energy difference, the stored elastic energy and the interfacial energy are all positive). Therefore, we have $A_s=A_f$.

In summary, in an ideally perfect single crystal and if only one martensitic plate is involved in the transformations, we have $M_s=M_f<T_0<A_s=A_f$. The forward and reverse transitions involve stored elastic energy but there is no thermoelasticity. The transitions are \textit{irreversible}.
\subsubsection*{Thermoelasticity with internal constraints} 
The situation is very different if the transitions involve some constraints that may impede the growth or the shrinkage of the martensitic domains, such as grain boundaries or pre-existing precipitates of a secondary phase. These constraints may be due also to the martensitic transition itself if many differently oriented martensitic plates are involved, or if some specific boundary conditions are imposed, such as a fixed volume in the presence of a dilatational component in the eigenstrain of the martensitic phase. 

As we explain below using two very simple generic models, these \textit{morphological constraints} may lead to a situation where a martensitic plate (if the constraints are \textit{local}, such as due to grain boundaries...) or the martensitic volume fraction (if the constraints are \textit{global}, such as fixed boundary conditions...) reach  a stable equilibrium state where the driving force due to the stress-free Gibbs free energy difference between the martensite and the austenite is exactly equilibrated by the driving force due to the stored elastic energy. In such state, the volume of the martensitic plate (or simply the martensitic volume fraction in the second situation) is proportional to $|\Delta g (T)|$, where  
$\Delta g (T)=g_m(T)-g_a(T)$ is the difference between the stress-free Gibbs free energy densities of the martensite and the austenite. In other words, in all cases, the volume fraction of martensite varies \textit{continuously} with the quantity of undercooling: this is precisely the definition of thermoelasticity. An important feature of this state is that the stored elastic energy consumes only a part of the available stress-free Gibbs free energy (exactly half in the two generic models presented below). If no other phenomena come into play (such as local free energy barriers due to the friction of the moving interfaces or due to the competition between differently oriented martensitic plates...), the martensitic volume fraction varies \textit{continuously} with temperature: the transition is reversible, even though stored elastic energy is involved. In other words, in its simplest expression, thermoelasticity by itself is a \textit{reversible} phenomenon. 

\subsubsection*{Irreversibility and intermittent dynamics}
However, the situation described above is oversimplified. In real systems, the moving interfaces encounter local free energy barriers of different nature. First, there is the presence of defects such as dislocations already present before the transformation or created during the process itself, either by partial plastic accommodation of the unrelaxed elastic energy along the habit planes or by the twinning mechanism itself. Second, there is the competitive growth between different martensitic plates. In all generality, the moving interfaces must overcome these defects: this is the origin of the pinning phenomena and the associated \textit{frictional work} . In term of the free energy landscape, these defects are associated to local energy barriers that cannot be overcome by thermal fluctuations. Therefore, for each event a finite amount of undercooling must be devoted to the decrease of the associated energy barrier \textit{until it disappears}. As soon as the barrier vanishes, the system becomes unstable and moves, at fixed temperature, until it meets the next barrier and becomes again pinned. From the thermodynamics point of view, the evolution at decreasing temperature between two consecutive jumps, that takes place at $T_1$ and $T_2<T_1$, is \textit{reversible}~\cite{endnote45} (the total entropy, i.e., the entropy of "universe",  stays constant) and the release of latent heat $\Delta Q$ during this process (i.e., measured by a calorimeter during a calorimetry experiment) is equal to the negative of the variation of enthalpy: $\Delta Q(T_1\rightarrow T_2)=H(T_1)-\hat H(T_2)$\cite{endnote46} (here and below, it is meant that the enthalpy $H$, and therefore the free enthalpy $G=H-TS$, includes the elastic energy). When temperature reaches $T_2$, the barrier collapses and the system jumps at fixed temperature to the next metastable state. This corresponding evolution is \textit{irreversible}, because the ending and initial states do not have the same Gibbs free energy: the entropy of universe increases by an amount $\Delta S_{tot}$ given by $\Delta S_{tot}(T_2)=-\frac{\Delta G(T_2)}{T_2}$, where $\Delta G(T_2)$ is the variation of the total Gibbs free energy. The jump at $T_2$ is of course also accompanied by a release of latent heat $\Delta Q (T_2)$ given by $\Delta Q(T_2)=\hat H(T_2)-H(T_2)$ (here and below, it is meant that the enthalpy $H$ and therefore the free energy enthalpy $G=H-TS$, includes the elastic energy). Part of this latent heat is released \textit{irreversibly}. 
This irreversible latent heat is precisely equal to the negative of the variation of the total Gibbs free energy during the jump: $\Delta Q_{irr}=-\Delta G(T_2)$. If the system was coupled to a mechanical device, the irreversible component of the released latent heat would correspond to "lost of useful work" due to irreversibility, i.e. to the part of the Gibbs free energy variation that is not converted into "useful work" (work other than PV).

In summary, thermoelasticity during the forward and reverse transformations can be observed if and only if these processes are accompanied by a finite unrelaxed stored elastic energy and if some internal (local) or morphological (global) constraints impede the growth or the shrinkage of the martensitic domains. Without local free energy barriers, such as those due to defects (usual "frictional work") or to the competing growth of different martensitic plates, the thermoelastic evolution would be thermodynamically reversible. In real systems, these barriers exist and the evolution consists of a series of small reversible quasi-static paths between consecutive irreversible jumps at fixed temperatures: \textit{thermoelasticity is then globally irreversible}.        

Such intermittent behavior  has  been observed in many slowly driven physical systems operating in nonequilibrium steady regimes associated, for instance, with magnetism, superconductivity, porous flow, friction, plasticity, fracture, earthquakes \cite{Fisher:1998fk,Papanikolaou:2011qf,PhysRevE.84.016115,Salman50}. These systems evolve toward a critical state represented by the ensemble of metastable states through which the system passes via an avalanche-like dynamics whose amplitudes and durations follow power laws (see Ref.  \cite{Sornette:2000ve} for a review). Similarly,  acoustic emission experiments as well as calorimetry measurements in temperature-driven athermal martensitic phase transformations have shown that the martensitic transition proceeds through avalanches  (see for example \cite{Vives48,Viv_Plan_1998,vives2010} , and also Ref.  \cite{gallardo} for an interesting discussion on the influence of the driving rate with respect to the role of an external disorder).

\subsubsection*{Model of Olson and Cohen}
The origin of thermoelastic behavior in martensites was first examined by Olson and Cohen \cite{Olson:1975rz} using a very simple thermodynamical model. In this model, the free energy change $(\Delta G)$ of a growing ellipsoidal poly-twinned martensitic particle of radius $r$ and thickness $c$ was expressed by
\begin{equation}
\label{eq:gener_olson}
\Delta G =  \frac{4}{3}\pi r^2c\Delta g_{ch} + 2\pi r c^2 \mu\epsilon_0^2+2\pi r^2\sigma, 
\end{equation}
where 
$\Delta g_{ch}(T)=g_m(T)-g_a(T)$ is the difference between the Gibbs free energy densities of stress-free bulk martensite and austenite, $\mu$ is a shear modulus, $\epsilon_0$ is a typical stress-free strain and $\sigma$ is the interfacial energy density. The first term at the r.h.s. of equation (\ref{eq:gener_olson}) corresponds to the stress-free driving force. It is supposed here~\cite{endnote47} that the martensitic particle  is polytwinned, in order to accommodate its bulk elastic energy. However, as already mentioned previously, this accommodation mechanism cannot be perfect. First, the boundary at the periphery of the ellipsoidal particle does not correspond to a habit plane. Therefore, this ribbon-like boundary generates a strain field which, if $c<<r$, is equivalent to the strain field generated by a dislocation. The corresponding unrelaxed elastic energy~\cite{endnote48} is given by the second term of equation (\ref{eq:gener_olson}). Second, the width of the twins along the habit planes is not infinitesimal and, consequently, there is a stray stress generated by the alternating domains. This stress is localized near the habit planes and thus the associated unrelaxed elastic energy is proportional to their areas. This contribution is included into the interfacial energy density $\sigma$, which in turn governs the last term of equation (\ref{eq:gener_olson}). 

\begin{figure}[ht!]
\begin{center}
\includegraphics[scale=0.45]{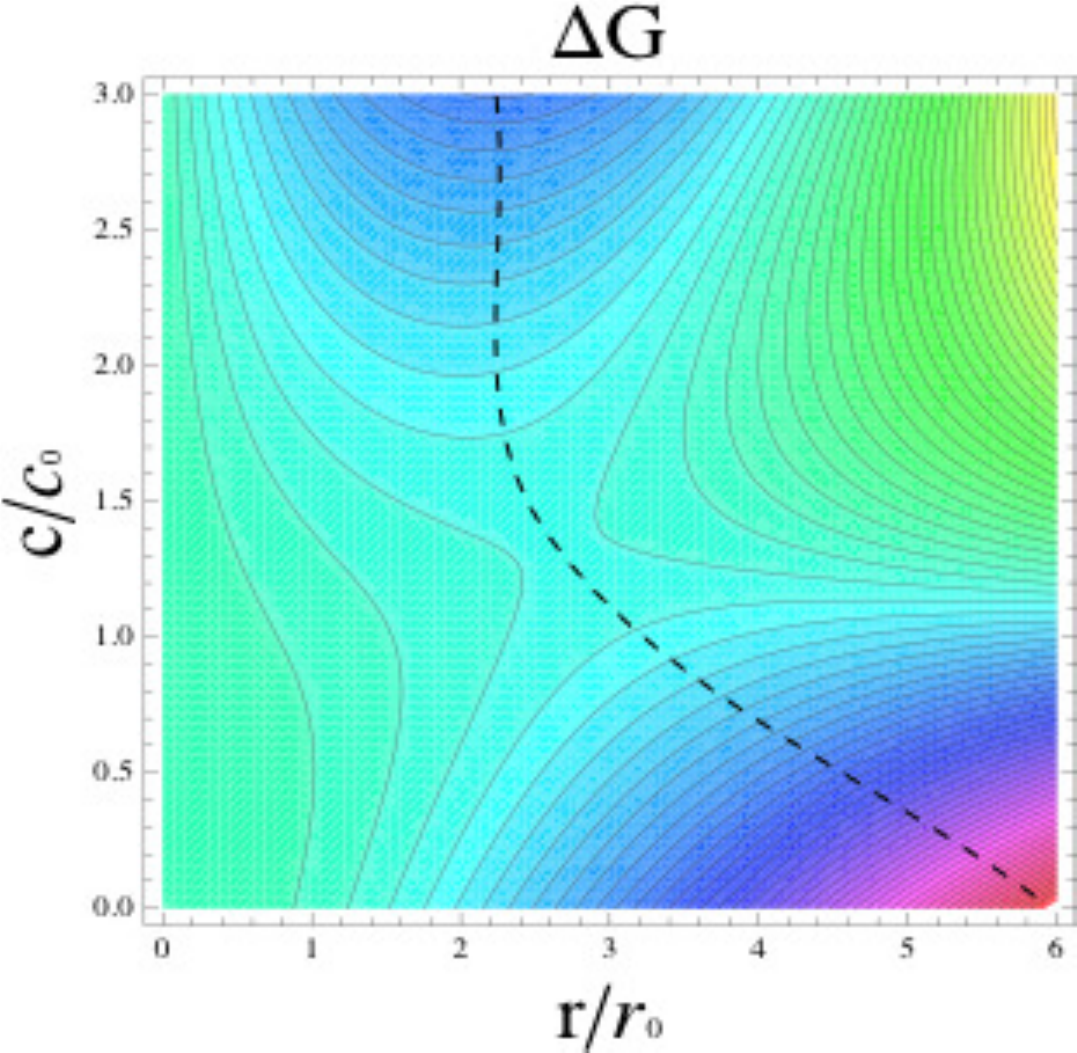}
\caption{\label{saddlepoint}\small\textit{Contour plot of the free energy density with respect to radius r and thickness c of the martensitic particle. The saddle point coordinates are given by: $r_{saddle}/r_0=\frac{8}{3}\,and\,c_{saddle}/c_0=\frac{4}{3}$, where  $r_0=\frac{9\mu\epsilon_0^2\sigma}{4\Delta g_{ch}(T)^2}$ and $c_{0}=-\frac{3\sigma}{2\Delta g_{ch}(T)}$.}}
\end{center}
\end{figure}

The contour plot of the free energy density is shown in Fig. \ref{saddlepoint}. It is seen that the stable and unstable regions are separated by the dashed line passing through  the  \textit{saddle point} with temperature dependent coordinates $r_s=\frac{6\mu\epsilon_0^2\sigma}{\Delta g_{ch}(T)^2}$ and $c_s=-\frac{2\sigma}{\Delta g_{ch}(T)}$.
At sufficiently small undercooling, the polytwinned domain is metastable, with finite radius and thickness. However, for a critical undercooling, the unstable region will move to the left due to temperature dependence. Therefore, the precipitate coordinates $r_{ini}$ and $c_{ini}$ will be in the  region (right side of the dashed curve shown in Fig. \ref{saddlepoint}) where the precipitate is unstable: the growth of a particle cannot be stopped in the absence of \textit{any morphological constraints such as grain boundaries, defects, fixed external boundaries or similarly other transformed domains of different orientations}. When radius and thickness of martensitic particle will be in the unstable region  after the required driving force is supplied, the martensite volume fraction will reach 100$\%$ at the transition start temperature without increasing the driving force. On the other hand, if the radial growth is stopped by any of these morphological constraints at a given radius $r=r_{max}$, the transformational force in the thickening direction leads to an energy balance condition between the stored strain energy and chemical driving force given by:
%
This lead to an equilibrium thickness given by:
\begin{equation}
\label{eq:c_eq}
c_{eq}(T) = \frac{-\Delta g_{ch}(T)}{3\mu\epsilon_0^2}r_{max}.   
\end{equation}
When this state is reached, the stored elastic energy $\Delta G_{el}=2\pi r_{max} c_{eq}^2 \mu\epsilon_0^2$ is exactly opposite to half of the available stress-free Gibbs energy $\Delta G_{ch}= \frac{4}{3}\pi r_{max}^2c_{eq}\Delta g_{ch}$:
\begin{equation}
\Delta G_{el}(T) = -\frac{\Delta G_{ch}(T)}{2}.
\end{equation}
As seen in equation (\ref{eq:c_eq}), the thickness of the martensitic particle beyond the critical undercooling varies continuously with temperature. This is the definition of the thermoelasticity. This simple model shows that the required condition for thermoelastic equilibrium is \textit{partial self-accommodation that leads to the stored elastic energy} and  \textit{internal constraints}.
\begin{figure}[ht!]
\begin{center}
\subfigure[][]{\label{fig:hysteresisthermoa}\includegraphics[scale=0.4]{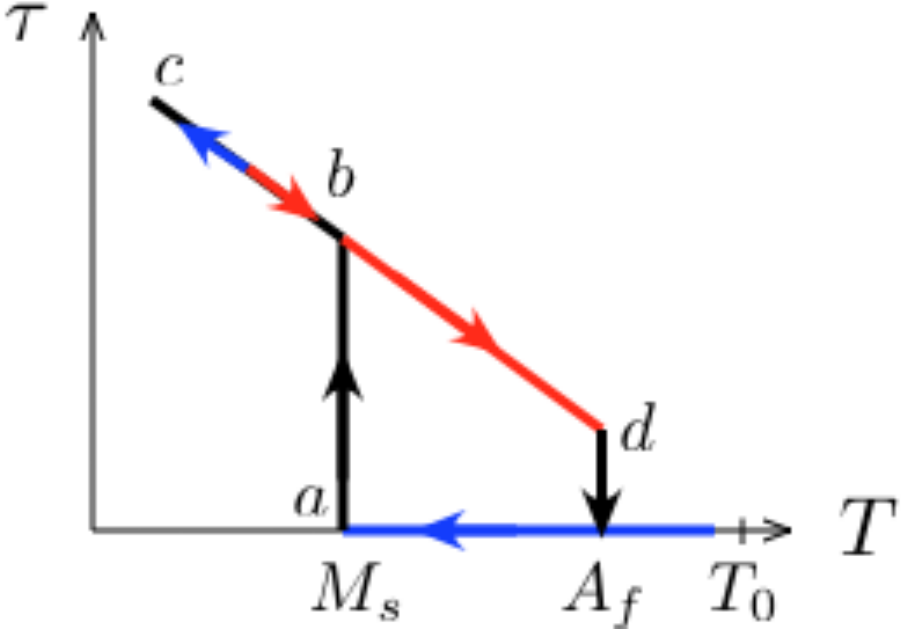}}\,
\subfigure[][]{\label{fig:hysteresisthermob}\includegraphics[scale=0.4]{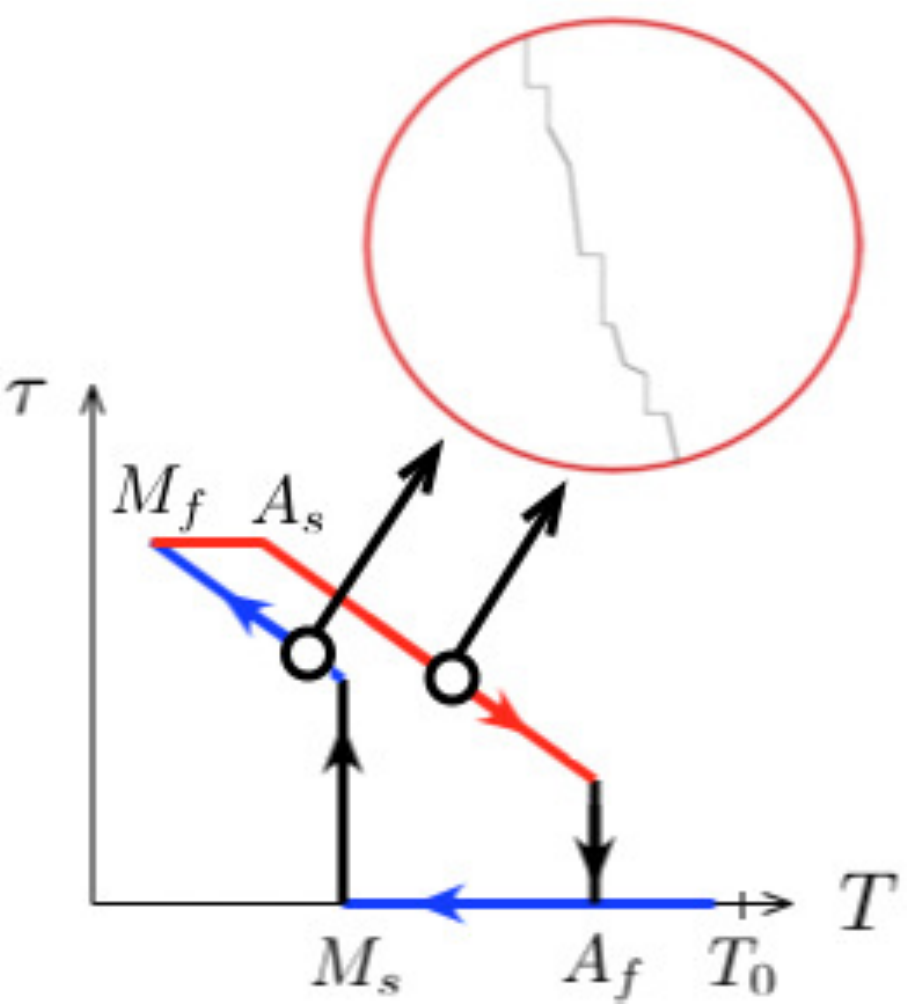}}\,
\caption{\label{fig:hysteresisthermo}\small\textit{Hysteresis of the growth of a single martensitic particle (a) in the absence of any kind of friction, (b) with friction. $\tau$ is the volume fraction of martensite and $T$ is the temperature.}}
\end{center}
\end{figure}
The corresponding hysteresis of the growth of a single martensitic particle is shown in Fig. \ref{fig:hysteresisthermo}. The transition starts at a martensite start temperature $M_s$ below the transition temperature $T_0$ defined by $\Delta g_{ch}=0$. This undercooling is required in order to overcome the energetic barrier due the interfacial energy and to the stored elastic energy, and in order to reach the saddle point. Afterwards, the martensitic particle grows along the segment (a-b) until its radial growth is stopped by a morphological constraint. At point b the particle reaches its thermoelastic equilibrium and along the segment (b-c) its growth is dictated by equation (\ref{eq:c_eq}), i.e., the transformation is frozen if the temperature is not decreased. On the other hand, when the system is heated, the martensitic particle retracts along the segment (c-d). It is easy to argue that, on heating, point d is beyond point b, i.e., $A_f>M_s$, and that $A_f$ is lower then the equilibrium temperature $T_0$. Indeed, on heating, the martensitic plate whose radius $r_{max}$ is imposed by some internal constraint, becomes unstable when the radius of the saddle point, which is temperature dependent, becomes equal to $r_{max}$. Obviously, the relation $r_{ini}<r_{max}<\infty$ implies $M_s<A_f<T_0$. Also, as the particle state at point d corresponds to an unstable state, the shrinkage at point d cannot be stopped. This scenario leads to the behavior displayed in Fig. \ref{fig:hysteresisthermo}\subref{fig:hysteresisthermoa}.     
    
The hysteresis described within this model (see Fig. \ref{fig:hysteresisthermo}\subref{fig:hysteresisthermoa}) does not take into account any frictional stress for interfacial motion  and it is reversible for temperatures below the critical undercooling, i.e., along segment (c-d) in Fig. \ref{fig:hysteresisthermo}\subref{fig:hysteresisthermoa}. However, the frictional forces are always present in a system (see the discussion above) and create local and small energetic barriers that must be overcome in order for the interfaces to move. This mechanism introduces metastable states in the system and therefore, the hysteresis shown in Fig. \ref{fig:hysteresisthermo}\subref{fig:hysteresisthermoa} becomes as shown in Fig. \ref{fig:hysteresisthermo}\subref{fig:hysteresisthermob}, which is  constituted by small bursts between these (meta)stable states, each of them at the thermoelastic equilibrium. 
\subsubsection*{Thermoelasticity for a non-volume preserving transformation and fixed boundary conditions}
In this subsection, we will show using a simple qualitative model that a non-volume preserving transformation together with fixed boundary conditions (zero macroscopic deformation or hard device boundary conditions) will lead to a thermoelastic equilibrium in the absence any other morphological constraints. 
    
To start with, we consider a system formed by the austenite and polytwinned martensitic domains with a non-zero dilatational part. We assume that the strain energy due to the shear components of the strain tensor is completely accommodated. But, of course, the dilatational part cannot be accommodated. The stress-free strains of martensite and the austenite are $\epsilon_0$ and 0, respectively. However, due to fixed boundary conditions, the actual strains $\epsilon_1$ and $\epsilon_2$ experienced by the martensite and the austenite will differ from the stress-free strains. We look here for a very qualitative picture. Therefore, as far as the elastic energy is concerned, we suppose that the strains $\epsilon_1$ and $\epsilon_2$  are homogenous. The total elastic energy $\Delta E_{strain}$ is then approximately given by:  
\begin{equation}
\frac{\Delta E_{strain}}{K V} =  \frac{V_1}{V}(\epsilon_1-\epsilon_0)^2 +\frac{V_2}{V}\epsilon_2^2,
\end{equation}
where $K$ is a bulk modulus, $V$ is the total volume of the system, and $V_1$ and $V_2$ the volume of the martensite and the austenite, respectively. For a macroscopic deformation imposed to zero, one can write 
\begin{equation}
\epsilon_1V_1+\epsilon_2V_2=0.
\end{equation} 
The above equation can also be rewritten in terms of volume fraction of each phase as
\begin{equation}
\epsilon_1\tau_1+\epsilon_2\tau_2=0;
\end{equation} 
where $\tau_1$ and  $\tau_2$  are the volume fractions of the martensite and the austenite, respectively. The strain energy in terms of the volume fractions becomes  
\begin{equation}
\label{eq:totalstrainmec}
\frac{\Delta E_{strain}}{K V} =  \tau_1(\epsilon_1-\epsilon_0)^2+\frac{\tau_1^2}{\tau_2}\epsilon_1^2,
\end{equation}
where we used $\epsilon_2=-\frac{\tau_1}{\tau_2}\epsilon_1$.
\begin{figure}[ht!]
\begin{center}
\includegraphics[scale=0.40]{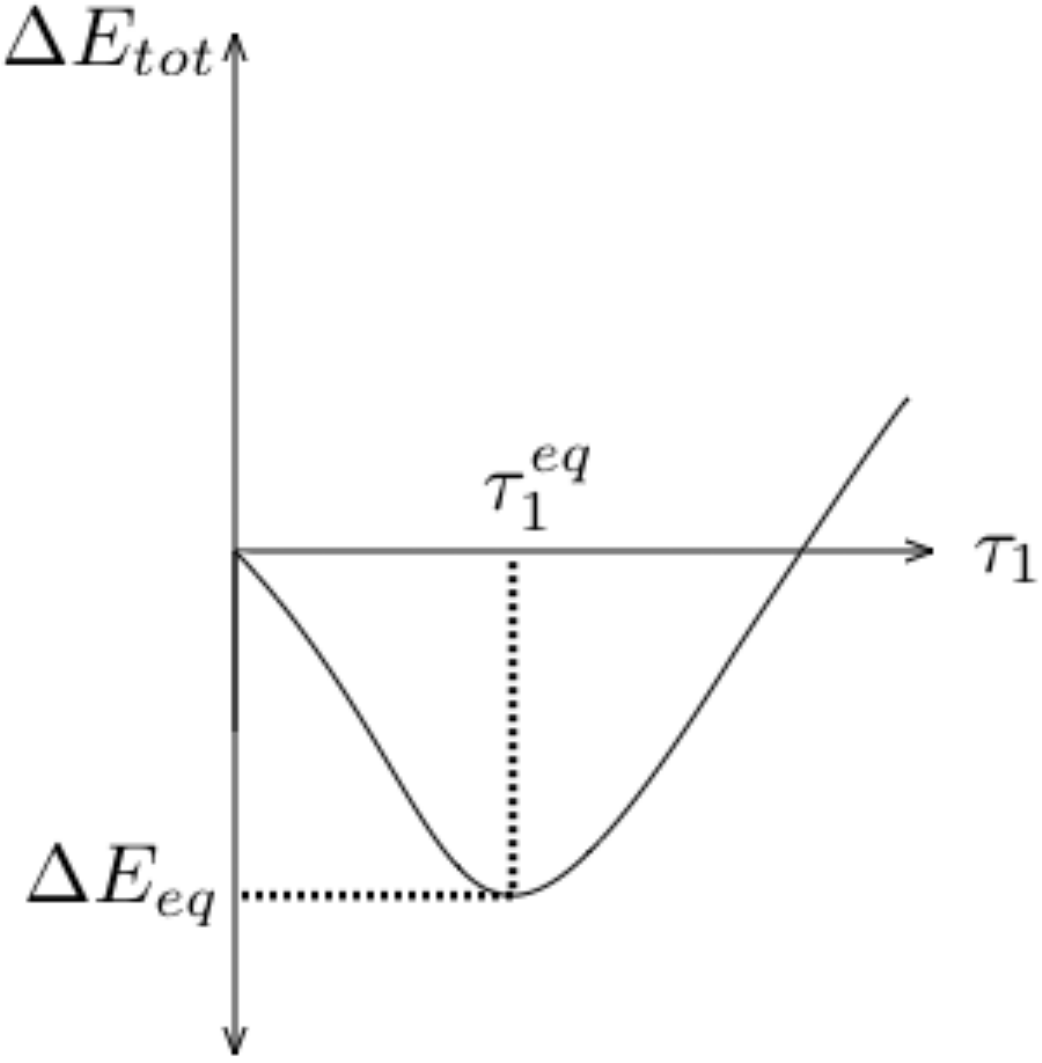}
\caption{\label{fig:volume}\small Total free energy as a function of martensitic volume fraction for a non-volume preserving transformation under clamped boundary conditions (zero macroscopic deformation).}
\end{center}
\end{figure}
The minimization of the above strain energy with respect to $\epsilon_1$ assuming that the strain energy quasi-statically  reaches its mechanical equilibrium expressed by $\frac{d\Delta E_{strain}}{d\epsilon_1}=0$, leads to    
\begin{equation}
\epsilon_1=\tau_2\epsilon_0.
\end{equation}
By substituting the above equation into equation (\ref{eq:totalstrainmec}), one finds
\begin{equation}
\label{eq:totalstrainres}
\frac{\Delta E_{strain}}{K V} = \tau_1^2\epsilon_0^2.
\end{equation}
The total free  energy change $\Delta E_{tot}$ including the stress-free Gibbs energy is then given by 
\begin{equation}
\label{eq:totalstrainchem}
\Delta E_{tot}=( \tau_1^2\epsilon_0^2K-\tau_1|\Delta g_{ch}|)V.
\end{equation}
This free energy change (see Fig.\ref{fig:volume}) has a minimum for 
\begin{equation}
\label{eq:c4_volfrac}
\tau_1^{eq} = \frac{|\Delta g_{ch}|}{2 K\epsilon_0^2}.
\end{equation} 
The above equation expresses the fact that for a given undercooling $|\Delta g_{ch}|$,  the martensite volume fraction will reach the value of $\tau_1^{eq}$. Afterwards, the volume fraction of martensite can only further increase (continuously) by supplying extra driving force. This is the definition of the \textit{thermoelastic equilibrium}. The value of the total free energy in the thermoelastic equilibrium is found to be  
\begin{equation}
\Delta E_{tot}^{eq}= -\frac{|\Delta g_{ch}|^2}{2 K\epsilon_0^2}V, 
\end{equation}
and again, as in the previous situation, the stored elastic energy consumes exactly half of the available chemical free energy:
\begin{equation}
\label{balance}
\Delta E_{strain}^{eq}= -\frac{1}{2}\tau_1^{eq}\Delta g_{ch}V = - \frac{1}{2} \Delta E_{ch}^{eq}
\end{equation}
where $\Delta E_{ch}^{eq}$ is the chemical free energy change associated to the formation of the martensitic volume fraction. 

For any further quasistatic temperature decrease $\delta T$, Eq. \ref{balance} indicates that the increase in stored elastic energy $\delta \Delta E_{strain}^{eq}$ always consumes exactly half the variation $\delta \Delta E_{ch}^{eq}$ of the chemical free energy made available by the increase of the martensitic volume fraction. In other words, only part of the available chemical free energy change is stored in the form of unrelaxed strain energy. However, eventhough the "internal work" stored in the transforming specimen is less than the available chemical free energy change, the thermoelastic transformation is thermodynamically reversible, simply because it consists in a succession of quasistatic equilibrium states. Because of the variational character of the total free energy at thermodynamic equilibrium, these states, correctly defined by the successive minimizations with respect to the strain and to the martensitic volume fraction in the present simplified model (or through the energy balance expressed in Eq. \ref{minimizationOC} in the previous example), correspond to true equilibrium states, and not to states where the equilibrium would only be established along the interfaces between the martensitic domains and the austenite, as suggested in \cite{Wollants:1993rz}. Also, further growth of the martensite is continuous and do not necessitate any finite increase of undercooling.

In conclusion, this simple model shows that in the absence of any other morphological constraints, \textit{the volume change associated with the martensitic transformation together with the macroscopic deformation imposed to zero} will cause the system to reach a \textit{reversible} thermoelastic equilibrium. 

\section{Conclusion}
In this study, microstructural evolutions of several  Ti-Ni-Pd and Ti-Ni-Cu ternary martensitic alloys have been investigated. For specific concentrations, these alloys satisfy a very non-generic condition such that the middle eigenvalue $\lambda_2$ of their transformation matrices is equal to 1. This condition allows the formation of compatible austenite/single variant of martensite interfaces. First, we realized a phase-field study on this subject and obtained different microstructural evolutions that depend on the value of $\lambda_2$. The microstructures obtained were in good agreement with the experiments, i.e., \textit{twinless domains are observed when $\lambda_2=1$, on the other hand  if $\lambda_2\neq1$, twinned martensite laminates appear.} However, we did not find significant differences between the transformation hysteresis of Ti$_{50}$Ni$_{27}$Pd$_{23}$ and Ti$_{50}$Ni$_{39}$Pd$_{11}$ alloys. Indeed, as it is explained above, the behavior of hysteresis in thermoelastic martensites depends, in principle, on frictional mechanisms. It is experimentally observed that when $\lambda_2=1$, the austenite/martensite interfaces are essentially coherent and display very few dislocations \cite{ISI:000274576500013}. This is probably due to the fact that, when $\lambda_2=1$, the compatibility between austenite and martensite may be fulfilled with a simple habit plane, i.e., without any transition layers associated to unrelaxed elastic energy. Generally, when $\lambda_2\neq1$, incompatibility enforces complex habit planes between polytwinned martensitic domains and the austenite, and it may be that the associated stored elastic energy is relaxed by plastic accommodation along the habit planes. Recently, this behavior was experimentally observed in reference \cite{Noreet:2009qr}.  As a result, it may be expected that internal friction of the moving interfaces is significantly higher when $\lambda_2\neq1$, resulting in a hysteresis larger than in the case $\lambda_2=1$.

In conclusion, the reason why we could not observe the differences in behavior of hysteresis of different alloys is understood by the fact that our phase-field model does not incorporate the dynamical dislocation creation during the transformation.  The relation of  the width of the transformation hysteresis to the value of $\lambda_2$ requires a modeling that couple the phase transformation with plastic activity  \cite{Levitas_1,Levitas_2}. Finally, our results suggest  that the behavior of hysteresis width cannot directly be understood by assuming that the condition $\lambda_2=1$  decreases only the critical nucleus size (see \cite{Zhang20094332}). It is probably a combined effect together with the decrease of  friction at the interfaces when $\lambda_2=1$. This conclusion has also been supported by  recent numerical studies \cite{Perez-Reche:2007et,Kastner_Eggeler_Weiss_Ackland_2011}, where the generation and  accumulation of persistent lattice defects during  forward-and-reverse phase transformations are observed.

%
\appendix
\section{}
The elastic   energy of  a coherent inclusion of a single variant of martensite (variant $p$) embedded into the austenite, can be calculated  using Eq.   \ref{eq:fambpq}      
\begin{equation}
\label{strain_inclusion_ch5}
\mathscr E_{elastic}=\frac{1}{2}\int_V^*B_{pp}(\bold{n})|\eta_p(\bold q)|^2\frac{\bold{dq}}{(2\pi)^3}.
\end{equation}
The matrices  $B_{pp}$ are calculated by inserting the deformation strain tensor $\epsilon_{ij}(p)$ of variant $p$ into Eq. \ref{eq:Bpq}. Since the term $B_{pp}(\bold{n})$ is positive, when the order parameter $\eta_p$ is different than $0$, the minimum of the strain energy is obtained for a platelet having a volume $V$ aligned in the direction of $\bold{n_0}$ that minimizes the function $B_{pp}(\bold{n})$.  We can numerically calculate the direction $\bold{n}_0$ that minimizes  the strain energy (\ref{strain_inclusion_ch5}). Calculations show that it exists a habit plane normal $\bold{n_0}$ for which the matrices $B_{pp}(\bold{n})$ vanish for the alloys Ti$_{50}$Ni$_{39}$Pd$_{11}$ and TiNiCu satisfying the condition $\lambda_2=1$. On the other hand, the matrices $B_{pp}(\bold{n})\neq0$ for any value of $\bold{n}$, when $\lambda_2\neq1$. It means that austenite/monovariant habit planes do not cost any strain energy only when $\lambda_2=1$.      

In order to calculate the elastic   energy of a coherent inclusion of a poly-twinned  domain, formed by two variants $p$ and $q$, embedded into the austenite,  we  calculate the average deformation of the poly-twinned martensite domain 
\begin{equation}
\bar\epsilon_{ij} = \kappa\epsilon_{ij}(p)  + (1-\kappa)\epsilon_{ij}(q),
\end{equation}
where $\kappa$ is the twin ratio. We can substitute the average deformation  into Eq. \ref{strain_inclusion_ch5} and minimize it with respect to $\kappa$, to  find the twin ratio and habit plane direction $n_0$ of the twinned-martensite domain with the austenite. We found that, when $\lambda_2=1$ condition is fulfilled,  the elastic energy approaches zero when twin ratio $\kappa$ goes to zero.
\bibliographystyle{tPHM}

\begin{thebibliography}{46}
\bibitem[41]{endnote43}{Note that we consider here that the elastic constants are homogeneous, i.e. they are same for the austenite and martensite phases.}
\bibitem[42]{endnote44}{From TEM, one never observes evidence of complete martensite structures in these modulated embryos in the austenite. The structures tend to evolve towards the martensitic structures, but as long as it is in the pre-martensite state, it has smaller deformations.}
\bibitem[43]{endnote45}{Provided of course that the cooling rate is small enough for the specimen to behave quasistatically.}
\bibitem[44]{endnote46}{The notation $\protect \mathaccentV {hat}05EH$ refer to a situation where the system is in a metastable state.}
\bibitem[46]{endnote48}{The energy of a dislocation loop that lies along the habit plane is given by $\sim \mu \protect \mathbf  b^2$, whose effective Burgers vector $\protect \mathbf  b$ is approximately $\protect \mathbf  b\sim \epsilon _0c$. The length of this effective dislocation loop is $2\pi r$. Therefore, its energy is simply given by $2\pi r \mu b^2\sim 2\pi rc^2\mu \epsilon _0^2$.}
\end{thebibliography}

\end{document}